\documentclass[aps,prx,twocolumn,showpacs,superscriptaddress,groupedaddress,floatfix]{revtex4-1}

\usepackage{amsmath}
\usepackage{amssymb}
\usepackage{graphicx}
\usepackage{multirow}
\usepackage{color} 
\usepackage{listings} 
\usepackage[title]{appendix}

\lstnewenvironment{alg}[1][]
{
\lstset{language={[95]Fortran},numberbychapter=true,frame=single,mathescape=true,commentstyle=\normalfont,basicstyle={\ttfamily},captionpos=b,linewidth=1.0\textwidth,xleftmargin=11pt,xrightmargin=11pt,framesep=11pt,escapechar=\%,float=*,#1 }
}
{}

\def\mch{d_0}
\def\ch{D}

\begin{document}

\title{Determining pressure-temperature phase diagrams of materials}

\author{Robert J. N. Baldock$^1$}
\email{rjnbaldock@gmail.com}
\affiliation{$^1$Cavendish, $^2$University Chemical and $^3$Engineering Laboratories, University of Cambridge, Cambridge, United Kingdom}
\author{L\'\i via B. P\'artay$^2$}
\affiliation{$^1$Cavendish, $^2$University Chemical and $^3$Engineering
Laboratories, University of Cambridge, Cambridge, United Kingdom}
\author{Albert P. Bart\'ok$^3$}
\affiliation{$^1$Cavendish, $^2$University Chemical and $^3$Engineering
Laboratories, University of Cambridge, Cambridge, United Kingdom}
\author{Michael C. Payne$^1$}
\affiliation{$^1$Cavendish, $^2$University Chemical and $^3$Engineering
Laboratories, University of Cambridge, Cambridge, United Kingdom}
\author{G\'abor Cs\'anyi$^3$}
\affiliation{$^1$Cavendish, $^2$University Chemical and $^3$Engineering
Laboratories, University of Cambridge, Cambridge, United Kingdom}

\date{\today}

\begin{abstract} 
We extend the nested sampling algorithm to simulate materials under periodic boundary  and  constant pressure conditions, and show how it can be  used  to determine the complete equilibrium phase diagram, for a given potential energy function, efficiently and in a highly automated fashion. The only inputs  required are the composition and the desired pressure and temperature ranges, in particular, solid-solid phase transitions are recovered without any{\em\ a priori} knowledge about the structure of solid phases. We benchmark and showcase the algorithm on the periodic Lennard-Jones system, aluminium and NiTi.

\end{abstract}

\maketitle

\section{Introduction}
\label{sec:intro}

 Phase diagrams of materials describe the  regions of stability  and equilibria of structurally distinct phases and are fundamental in both materials science and industry. In order to augment experiments, computer simulations and theoretical calculations are often used to provide reference data and describe phase transitions. 
A plethora of methods exist to determine individual phase boundaries, including Gibbs ensemble Monte Carlo~\cite{GEMC}, Gibbs--Duhem integration~\cite{G-D_int}, thermodynamic integration and even direct molecular dynamics simulations of coexistence.
Each of these algorithms requires the user to specify at least the identity and approximate location of the phase transition under investigation.
Moreover, in the case of the solid phases, where much of the interest lies, advance knowledge of the crystal structure of each  phase is required.
Calculating an entire phase diagram by combining the results of such methods therefore demands a high degree of prior knowledge of the result. 
This in turn poses a barrier to the discovery of unexpected phases and phase transitions.
Furthermore, such algorithms require specific expertise and separate setup for each type of phase transition.

In this paper we introduce a single algorithm, based on nested sampling (NS)~\cite{bib:skilling,bib:skilling2}, that enables the efficient calculation of {\em complete} pressure-temperature phase diagrams, including the solid region. 
This algorithm requires no prior knowledge of the phase diagram, and takes only the potential energy function together with the desired pressure and temperature ranges as inputs.
Moreover, the direct output of the simulation is the partition function as an explicit function of its natural variables, so calculating  thermodynamic observables, such as the heat capacity, is straightforward.

Nested sampling systematically explores the entire potential energy landscape, and in this way is related to parallel tempering (also known as replica exchange)~\cite{bib:partemp_swendsen,bib:partemp} and Wang-Landau sampling~\cite{bib:wang_landau}.
However, those algorithms encounter a particular convergence problem at first order phase transitions 
because
the probability distributions (parametrised in terms of temperature in case of
 parallel tempering or energy in case of Wang-Landau) on the two sides of the phase transition have very little overlap~\cite{AllenTildesley}. 
 This results
 in poor 
 equilibration between the distributions on either side of the phase transition
 and large errors (both random and systematic) in the predicted locations of phase transitions.

The NS algorithm was designed to solve this problem. 
It constructs a sequence of decreasing potential energy levels, $\{E_i\}$, each of which bounds from above a volume of configuration space $\chi_i$,  with the property that $\chi_i$ is approximately a constant factor smaller than the volume, $\chi_{i-1}$, corresponding to the level above. Each volume is sampled uniformly, and therefore each distribution will have an approximately constant fractional overlap with the one immediately before and after, ensuring fast convergence of the sampling and allowing an accurate evaluation of phase space integrals. In particular, the energy levels near the phase transition, where phase volumes change rapidly, will be very narrowly spaced. The sequence of energy levels comprise a discretisation of the cumulative density of states $\chi(E)$, which allows the evaluation of the partition function at arbitrary temperatures,
\begin{eqnarray}
Z(N,V,\beta) &=& \frac1{N!}\left(\frac{2\pi m}{\beta h^2}\right)^{3N/2}\int dE  \,\chi'(E) e^{-\beta E} \label{eq:Z_NVT}\\
&\approx&Z_m(N,\beta)\sum_i (\chi_{i-1}-\chi_i) e^{-\beta E_i}\label{eq:Z_NS}
\end{eqnarray}
where $N$ is the number of particles of mass $m$, $V$ is the volume, $\beta$ is the inverse temperature, $h$ is Planck's constant,  the density of
states $\chi'$ is the  derivative of $\chi$, and we labelled the factor resulting from the momentum integral as $Z_m$. The total phase space volume is  $\chi_0=V^N$ corresponding to the ideal gas limit. Note that the  sequence of energies $\{E_i\}$ and configuration space volumes $\{\chi_i\}$  are independent of temperature, so the partition function can
be evaluated {\em a posteriori} at any  temperature by  changing $\beta$ in \eqref{eq:Z_NS}.

The basic NS algorithm is as follows. We  initialise by generating a pool of   $K$ uniformly random configurations and   iterate the following loop starting at $i=1$.

\begin{enumerate}
        \item Record the energy of the sample with the highest energy as $ E_{i} $, and use it as the new energy limit, $E_{\mathrm{limit}} \leftarrow E_i$. The corresponding phase space volume is  $\chi_{i}\approx \chi_0 [K/(K+1)]^i$.
        \item Remove the sample with energy $E_i$ from the pool and generate a new configuration uniformly random in the configuration space, subject to the constraint that  its  energy is less than $E_{\mathrm{limit}}$. One way to do this is to clone a randomly chosen existing configuration and make it undergo a random walk of $L$ steps,  subject only to the  energy limit constraint.
        \item Let $i \gets i +1$, and return to step 1.
\end{enumerate}

At each iteration, the pool of  $K$ samples are uniformly distributed in configuration space with energy  $E<E_{\mathrm{limit}}$. The finite sample size leads to a  statistical error in $\log \chi_i$, and also in the computed observables, that is asymptotically proportional to $1/\sqrt{K}$, so any desired accuracy can be achieved by increasing $K$. Note that for any given $K$, the sequence of energies and phase volumes converge exponentially fast (the number of
iterations required to obtain results shown below never exceeded   $2000\cdot
K$), and  increasing $K$ necessitates a new simulation from scratch.  

Since its inception NS has been used successfully for Bayesian model selection in astrophysics~\cite{multinest1}, and also to investigate the potential energy landscapes of atomistic systems ranging from clusters to proteins~\cite{bib:our_NS_paper,bib:our_NSHS_paper,Burkoff1,bib:wheatley1,bib:wheatley2,bib:wheatley3, neilsen_npt,  bib:Frenkel_NS,bib:diffns}.

The structure of this paper is as follows.
In section~\ref{sec:npt_ns}  we modify the NS algorithm to enable its application at constant isotropic pressure with fully flexible periodic boundary conditions~\cite{martyna1994constant} where the periodic simulation cell is allowed to change shape.
In sections~\ref{sec:calc_phase_dia} and~\ref{sec:results} we show that this development enables the determination of pressure-temperature phase diagrams of materials directly from the potential energy function without recourse to any other{\em\ a priori} knowledge.
In particular, in section~\ref{sec:results} we calculate phase diagrams for aluminium and NiTi.
Finally in section~\ref{sec:conc_outlook} we conclude this paper, discussing some consequences of the capability to calculate entire phase diagrams with a single method and in a highly automated fashion.

\section{Nested sampling with fully flexible periodic boundary conditions at constant pressure}
\label{sec:npt_ns}

Nested sampling produces new samples by cloning an existing sample and then evolving the clone using a Markov chain Monte Carlo (MCMC) random walk~\cite{brooks2011handbook}. Although one could work in the NVT ensemble and use equations~\ref{eq:Z_NVT} and \ref{eq:Z_NS}, that would be very inefficient. 
MCMC simulations performed at fixed pressure require just a fraction of the computational expense as equivalent calculations performed at fixed volume.
There are two reasons for this.

First, allowing the system to change volume by dilating or contracting expedites the cooperative freeing of jammed atoms.
In contrast, at fixed volume, atoms that have become jammed are only freed by the coincidental movement of all atoms to separate them.
Consequently MCMC simulations at fixed pressure explore configuration space far more rapidly than simulations at fixed volume.

The second reason arises from the thermodynamic behaviour of systems at a first order phase transition.
At a phase transition under constant volume conditions the two phases coexist and an interface forms between them.
Such interfaces are large on the atomic scale~\cite{frenkel2001understanding}
and 
the behaviour of atoms at an interface is not representative of the behaviour of atoms in the equilibrium phases. 
As a result the interface introduces a systematic error that is only overcome by simulating very large numbers of atoms.

Such interfaces also occur under constant pressure conditions in the infinite system size limit.
The contribution to the Gibbs Free Energy from an interface is proportional to $\gamma N^{\frac{2}{3}}$, where $\gamma$ is the interfacial tension.
In contrast, the Gibbs Free Energies of each of the pure phases are extensive (proportional to $N$).
Therefore the Gibbs Free Energy cost of the interface is negligible for thermodynamic systems.
Conversely, for the relatively small system sizes amenable to density of states calculation methods such as nested sampling, the Gibbs Free Energy cost of the interface is appreciable, provided $\gamma$ is not close to zero.
Consequently, at a constant pressure phase transition between phases with identical atomic compositions, configurations containing an interface have negligible statistical weight in such simulations, and a discontinuous transition is observed from one equilibrium phase to the other.
This enables the accurate simulation of phase transitions using much smaller numbers of atoms.

Using small numbers of atoms to simulate a phase transition naturally introduces new finite size errors.
In particular, for a fixed number of atoms, it is not possible to represent all crystal structures in a simulation cell of fixed shape.
This representational bias is removed by the use of fully flexible periodic boundary conditions~\cite{martyna1994constant}, which allow the simulation cell to deform smoothly and thus take any shape.
However, using fully flexible periodic boundary conditions allows the formation of very thin simulation cells containing unphysical quasi one and two dimensional configurations, characterised by interacting periodic images.
In subsection~\ref{subsec:mindepth_theory} we describe a rigorous solution to this new finite size problem.
Later, in subsection~\ref{subsec:part_fn_and_thermodynamics} we describe the calculation of the constant pressure partition function and heat capacity, both as explicit functions of temperature, using nested sampling.

\subsection{Constraint on the simulation cell to exclude unphysical quasi one and two dimensional configurations}\label{subsec:mindepth_theory}

The partition function at fixed isotropic pressure $p$ with fully flexible periodic boundary conditions~\cite{martyna1994constant} is
\begin{equation}
\begin{aligned}
\Delta(N,p,\beta) & = Z_m  \beta p  \int d\mathbf{h}_{0}  \delta\left(\det{\mathbf{h}_{0}}-1\right) \times \\ 
& \int_{0}^{\infty} \!\!d V V^N \int_{\left(0,1\right)^{3N}}\! d\mathbf{s} \, e^{-\beta H\left( \mathbf{s},\mathbf{h}_{0},V,p \right)}.
\end{aligned}
\label{eq:partition_fn_NPT}
\end{equation}
Here $H\left( \mathbf{s},\mathbf{h}_{0},V,p \right) = E\left( \mathbf{s},\mathbf{h}_{0},V \right) + pV $, $\mathbf{h}$ is the  $3\times3$  matrix of lattice vectors relating the Cartesian positions of the atoms $\mathbf{r}$ to the fractional coordinates $\mathbf{s}$ via  $\mathbf{r}=\mathbf{hs}$, $ V=\det\mathbf{h}$ is the volume, and $\mathbf{h_{0}}=\mathbf{h}V^{-1/3}$ is the image of the unit cell normalised to unit volume.

The partition function~\eqref{eq:partition_fn_NPT} corresponds to integration over all nine elements of the matrix $\mathbf{h}_{0}$, and the $\delta$-function restricts the integration to matrices satisfying $ \det\mathbf{h}_{0} = 1$.
This partition function is  formally correct in the thermodynamic limit~\cite{martyna1994constant,tuckerman2008statistical}.
However, finite systems in this  description can adopt configurations for which the simulation cell becomes very thin.
In this case, periodic boundary conditions give rise to a quasi one or two dimensional system.
The prevalence of such configurations leads to a poor approximation of the three dimensional atomic system due to excessively large finite size effects.
We exclude such thin configurations by changing the limits for integration over elements of $ \mathbf{h}_{0}$, so that the perpendicular distances between opposite faces of the simulation cell $ \mathbf{h}_{0}$ are greater than some ``minimum cell depth'' value $\mch$.

The perpendicular distance between faces of the unit cell $\mathbf{h}$  made by lattice vectors $ \mathbf{h}^{\left( i \right)} $ and  $ \mathbf{h}^{\left( j \right)} $ is given by
\begin{equation} \label{eq:d_perp}
  d^{\perp}_{\mathbf{h}^{\left( k \right)}} = \frac{  \det \mathbf{h}  }{| \mathbf{h}^{\left( i \right)}\times\mathbf{h}^{\left( j \right)} | }.
\end{equation}
The cell depth $\ch\!\left(\mathbf{h}_0\right)$, which measures how ``thin'' the cell has become, is defined as the minimum value of $d^{\perp}_{\mathbf{h}^{\left( k \right)}}$, for the cell at normalised (unit) volume $\mathbf{h}_0$.
\begin{equation} \label{eq:cell_depth}
\ch\!\left(\mathbf{h}_0\right) = \min_{i=1,2,3}\left(d^{\perp}_{\mathbf{h}_0^{\left( i \right)}} \right)
\end{equation}
Thus we integrate over elements of $ \mathbf{h}_{0}$ such that 
\begin{equation} \label{eq:min_cell_depth_crit}
\ch\left(\mathbf{h}_0\right) > \mch.
\end{equation}
The minimum cell depth $\mch$ is a real number on the interval $[0,1]$ where $\mch=1$ restricts the simulation cell to a cube.
Smaller values of $\mch$ are accordingly less restrictive on the shape of the simulation cell, and $\mch=0$ corresponds to no restrictions on the simulation cell.

Incorporating this change of integration limits into the partition function~\eqref{eq:partition_fn_NPT} yields a new partition function
\begin{equation}
\begin{aligned}
\widetilde{\Delta}(N,p,\beta,\mch) & = Z_m  \beta p  \int_{\ch\left(\mathbf{h}_0\right) > \mch} \!\!\!\!\!\!\!\!\! d\mathbf{h}_{0}  \delta\left(\det{\mathbf{h}_{0}}-1\right) \times \\
& \int_{0}^{\infty} \!\!d V V^N  \int_{\left(0,1\right)^{3N}}\! d\mathbf{s} \, e^{-\beta H\left( \mathbf{s},\mathbf{h}_{0},V,p \right)   }.
\end{aligned}
\label{eq:partition_fn_NPT_w_mch}
\end{equation}
In the thermodynamic limit~\eqref{eq:partition_fn_NPT_w_mch} is equal to~\eqref{eq:partition_fn_NPT} up to a factor which depends only on  $\mch$.
The two partition functions are equal if and only if $\mch=0$.

In tests with 64 atoms we verified that the heat capacity curves were independent of $\mch$ at values of 0.65, 0.7 and 0.8, in Lennard-Jonesium and aluminium. 
The window of independence from $\mch$ grows wider as the number of particles is increased. 
For larger numbers of atoms, there are more ways to arrange those atoms into a given crystal structure, including in simulation cells that are closer to a cube.
Similarly, unphysical correlations are introduced when the \emph{absolute} number of atoms between faces of the cell becomes too small, and therefore larger simulations can tolerate ``thinner'' simulation cells $\mathbf{h}_{0}$.
The nickel-titanium calculations were performed with $\mch=0.7$.

\subsection{Partition function and thermodynamic variables}\label{subsec:part_fn_and_thermodynamics}

The partition function we seek to calculate is given in equation~\eqref{eq:partition_fn_NPT_w_mch}.
Above some sufficiently large volume $V_0$, we approximate the system as an ideal gas, neglecting interatomic interactions, which corresponds to the condition $E\left( \mathbf{s},\mathbf{h}_{0},V \right) \ll pV$.
In this approximation the volume integral in~\eqref{eq:partition_fn_NPT_w_mch} is the sum of two parts
\begin{equation}
\begin{aligned}
 & \widetilde{\Delta} (N,p,\beta,\mch)  \approx Z_m  \beta p  \Bigg[ \Delta_{\mathrm{NS}} (N,p,\beta,V_0,\mch)   \\
& + \int_{\ch\left(\mathbf{h}_0\right) > \mch} \!\!\!\!\!\!\!\!\! d\mathbf{h}_{0}  \delta\left(\det{\mathbf{h}_{0}}-1\right) \int_{V_0}^{\infty} \!\!d V V^N  \int_{\left(0,1\right)^{3N}} \! d\mathbf{s} \, e^{-\beta pV } \Bigg]
\end{aligned}
\label{eq:int_split}
\end{equation}
where
\begin{equation}
\begin{aligned}
 \Delta_{\mathrm{NS}} (N,p,\beta,V_0,\mch) & = \int_{\ch\left(\mathbf{h}_0\right) > \mch} \!\!\!\!\!\!\!\!\! d\mathbf{h}_{0}  \delta\left(\det{\mathbf{h}_{0}}-1\right) \times \\
& \int_{0}^{V_0} \!\!d V V^N \int_{\left(0,1\right)^{3N}}\! d\mathbf{s} \, e^{-\beta\left[  E\left( \mathbf{s},\mathbf{h}_{0},V \right) + pV \right]} 
\end{aligned}
\label{eq:pf_NS}
\end{equation}

We calculate $\Delta_{\mathrm{NS}}$ using nested sampling.
Calculations are performed at fixed pressure to generate a sequence of {\em enthalpies}, $H_i$, where  $H=E\left( \mathbf{s},V,\mathbf{h}_{0}\right) + pV$.
The NS approximation for $\Delta_{\mathrm{NS}}$, is
\begin{equation}\label{eq:DD_NS}
\begin{split}
\Delta_{\mathrm{NS}} (N,p,\beta,V_0,\mch) \approx& \sum_{i=1}^{i_{\mathrm{max}}}{\left(\chi_{i-1}-\chi_{i}\right) e^{-\beta H_i}  } \\
\approx& \sum_{i=1}^{i_{\mathrm{max}}}{\Delta\chi_i e^{-\beta H_i}  }
\end{split}
\end{equation}
where  $\chi_{i} \approx \chi_0\left(\frac{K}{K+1}\right)^i$, $\chi_0 = \frac{V_0^{N+1}}{N+1}$, and  $\Delta\chi_i \approx \chi_{i-1}-\chi_{i}$.
We use single atom Monte Carlo (MC) moves  in fractional coordinates with the amplitude updated every $\frac{K}{2}$  iterations to maintain a good acceptance rate. 
Uniform sampling of lattice shape matrices $\mathbf{h}_{0}$ subject to equation~\eqref{eq:min_cell_depth_crit} was achieved by independent shearing and stretching moves which do not change the volume.
The ratios of atom, volume, shear and stretch moves were $N\!\!:\!\!10\!\!:\!\!1\!\!:\!\!1$. 
Further details of the MC moves and parallelisation scheme are given in the Supplemental Material (SM)~\cite{SM}.

We show in appendix~\ref{app:ig_pf} that volumes greater than $V_0$ make a negligible contribution to the partition function~\eqref{eq:int_split}, provided $ k_{\mathrm{B}}T \ll  pV_0 $.
In this case we have
\begin{equation} \label{eq:pf_notail}
\widetilde{\Delta}(N,p,\beta,\mch)  \approx \frac {  \beta p  } {N!} \Biggl( \frac {2\pi m } {\beta h^2}\Biggr)^{3N/2}  \Delta_{\mathrm{NS}}(N,p,\beta,V_0,\mch)
\end{equation}
where we have expanded $Z_m$.
One can always assert the condition $ k_{\mathrm{B}}T \ll  pV_0 $, and in practice  it is easy to find values of $V_0$ suitable for physically relevant conditions.
We found $V_0 = 10^7 N \: \mathrm{\AA^3}$ to be suitable for all conditions considered in this paper.
From~\eqref{eq:pf_notail} we obtain the expected enthalpy
\begin{equation} \label{eq:expected_H}
\begin{split}
\langle H \rangle =& -\frac {\partial \log \widetilde{\Delta}( N , p ,\beta,\mch)} {\partial \beta} \\
=& \left( \frac{3 N }{2} - 1 \right)\frac{1}{\beta} +   \langle H_{\mathrm{configurations}} \rangle  
\end{split}
\end{equation}
and the heat capacity at constant pressure
\begin{align}
C_p =& - k_{\mathrm{B}}\beta^2 \frac {\partial \langle H \rangle } {\partial \beta} \label{eq:cp_line1} \\
=& \left(\frac{3 N   }{2}-1\right)k_{\mathrm{B}} \label{eq:cp_line2}   \\ 
 & + { k_{\mathrm{B}}\beta^2} \left({\langle H^2_{\mathrm{configurations}} \rangle  - { \langle H_{\mathrm{configurations}} \rangle}^2 }\right) \nonumber
\end{align}
where
\begin{equation} \label{eq::moments_tail_cont_zero}
\begin{split}
\langle H_{\mathrm{configurations}}\rangle &\approx  \frac{ \sum_{i=1}^{i_{\mathrm{max}}}{ \Delta\chi_i \:  H_i \: e^{-\beta H_i}}  }{ \sum_{i=1}^{i_{\mathrm{max}}}{ \Delta\chi_i \: e^{-\beta H_i}}    },  \\
\langle H^2_{\mathrm{configurations}}\rangle &\approx  \frac{ \sum_{i=1}^{i_{\mathrm{max}}}{ \Delta\chi_i \:  H^2_i \: e^{-\beta H_i}}  }{ \sum_{i=1}^{i_{\mathrm{max}}}{ \Delta\chi_i \: e^{-\beta H_i}}    }.
\end{split}
\end{equation}
This form~\eqref{eq::moments_tail_cont_zero} naturally does not depend on the contribution made by the low density configurations omitted from the NS calculation, or explicitly on the value of $\mch$.
We used equations~\eqref{eq::moments_tail_cont_zero} when calculating the heat capacities presented in this paper.

\section{Calculating phase diagrams}
\label{sec:calc_phase_dia}

In this section we describe a method for calculating the phase diagram of a material from the output of nested sampling.
We then benchmark the performance of nested sampling on the periodic Lennard-Jones system, and find nested sampling to be orders of magnitude more efficient than Parallel Tempering (PT) for resolving the melting and evaporation transitions.

\begin{figure}
\begin{center}
\includegraphics[width=8.5cm]{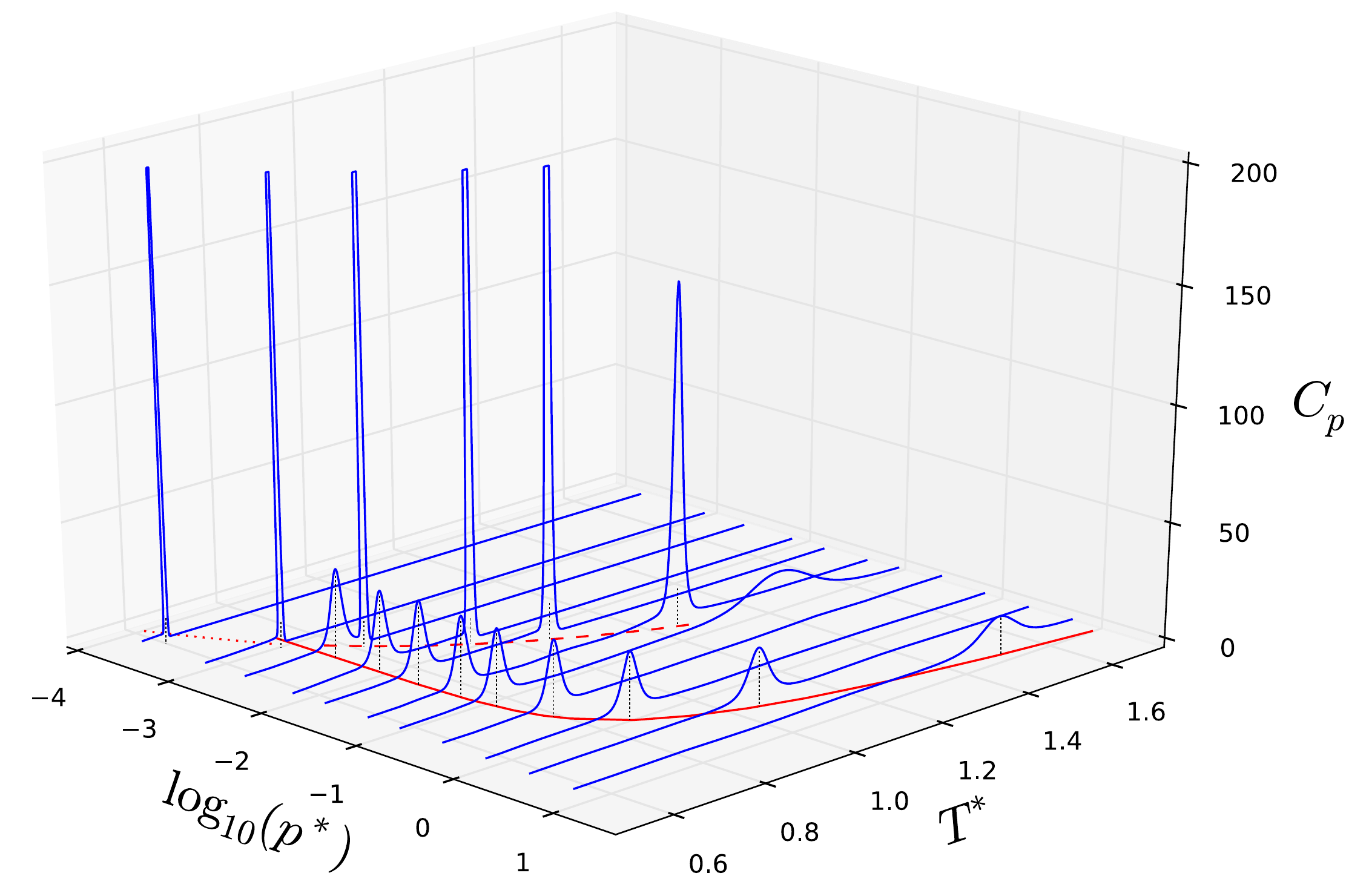}
\end{center}
\vspace{-15pt}
\caption {Demonstration of how NS can be used to calculate phase diagrams, using the case of the periodic Lennard-Jones model.
NS calculations are performed at a series of pressures and phase transitions are located by peaks of the heat capacity curves (blue).
The red lines show values from the literature for the melting (solid)~\cite{McNeilWatson2006}, boiling (dashed)~\cite{Kofke1993} and sublimation (dotted)~\cite{AgrawalKofke1995} curves.}
\label{fig:lj_phase_dia_from_cps}
\end{figure}

Given the partition function~\eqref{eq:pf_notail}, phase transitions can be easily located by finding the peaks of response functions such as the heat capacity~\eqref{eq:cp_line2}.
By performing separate NS simulations at a number of pressures and combining the pressure and temperature values corresponding to the heat capacity peaks  one can straightforwardly construct the \emph{entire}  phase diagram including all thermodynamically stable phases.
This process is illustrated in Figure~\ref{fig:lj_phase_dia_from_cps}.

\begin{figure}
\begin{center}
\includegraphics[width=8.5cm]{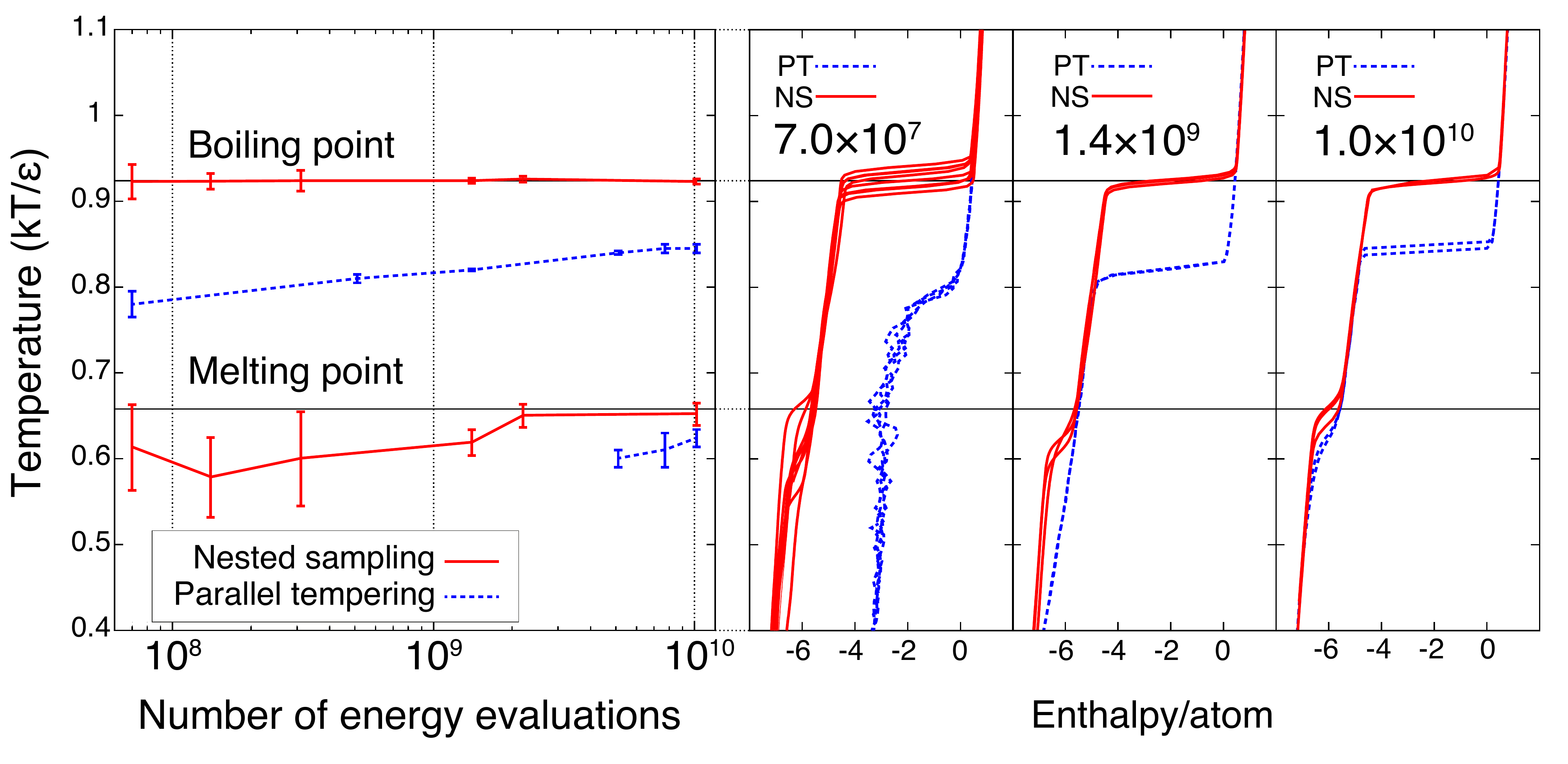}
\end{center}
\vspace{-15pt}
\caption {Performance comparison of NS and PT.
Sixty-four Lennard-Jones particles were simulated at a pressure of $0.027$ (Lennard-Jones units).
Both NS and PT simulations were initialised from the vapour phase. 
PT was performed using 128 equispaced temperature values in the range $[0.4,1.4]$. 
The left panel shows the estimated transition temperatures as a function of computational cost while the right panel shows the mean enthalpy as a function of temperature corresponding to three selected values of the cost.
} 
\label{fig:lj_nspt}
\end{figure}

In Figure~\ref{fig:lj_nspt} we compare the performance of NS to that of PT for calculating the melting and evaporation transitions.
NS provides a reasonable estimate of the melting and boiling points using only $\sim 10^{8}$ energy evaluations, while parallel tempering needs many orders of magnitude more computational effort than NS to find the evaporation transition and almost two orders of magnitude more computational effort to find the melting transition. (A similar increase in computational efficiency  compared with parallel tempering was found for LJ clusters~\cite{bib:our_NS_paper} and  hard spheres~\cite{bib:our_NSHS_paper,bib:HS_PT}.)

\begin{figure}
\begin{center}
\includegraphics[width=8.5cm]{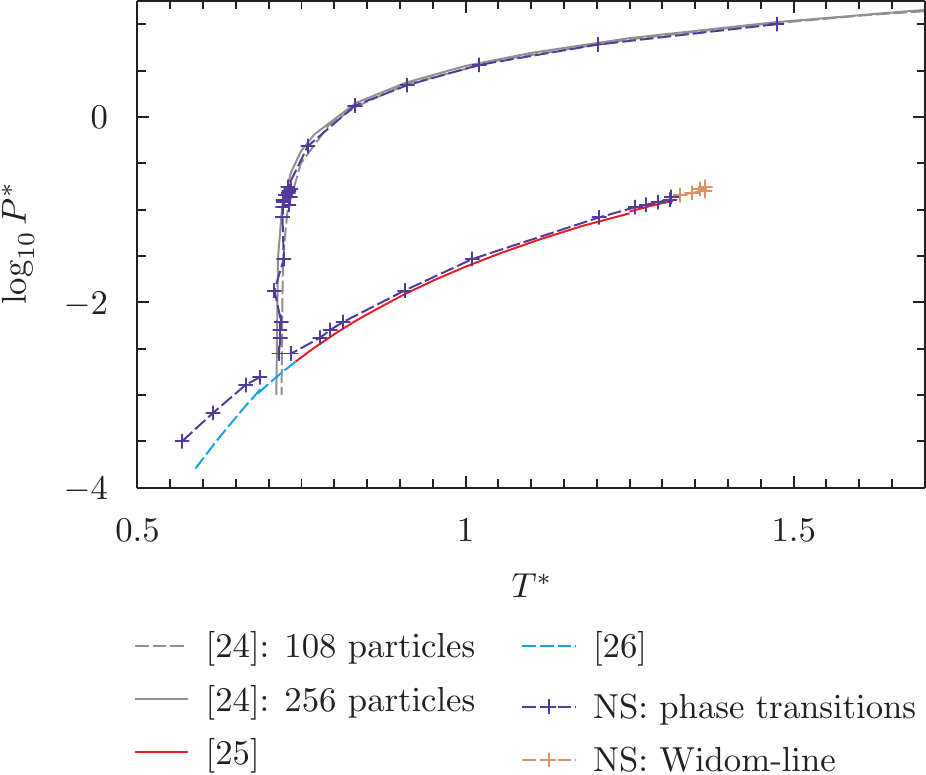}
\end{center}
\vspace{-15pt}
\caption {Phase diagram for $N=64$ Lennard-Jones particles as calculated using NS, with comparison to the literature ($N\approx500$) phase diagram, as described in the text.
} 
\label{fig:lj_pd}
\end{figure}

Finally, in Figure~\ref{fig:lj_pd} we show the phase diagram for 64 particles of Lennard Jonesium as calculated using NS with $K=640$, $L=1.6\times10^5$.
Comparison with the literature phase diagrams for $\sim 500$ particles confirms excellent agreement with the literature values for the evaporation transition~\cite{Kofke1993} and also the solid-liquid and high pressure solid-vapour transitions~\cite{McNeilWatson2006}.
Below the triple point, we observe slower convergence with respect to $L$ towards literature values of the sublimation transition~\cite{AgrawalKofke1995}.
We also find the beginning of the Widom-line: the shallow line of heat capacity maxima that extends into the supercritical region.
The Widom-line and our method for estimating the critical point are described in the SM~\cite{SM}.

\section{Results}\label{sec:results}
\subsection{Aluminium}\label{subsec:al}

\begin{figure*}[ht]
\includegraphics[width=17.5cm]{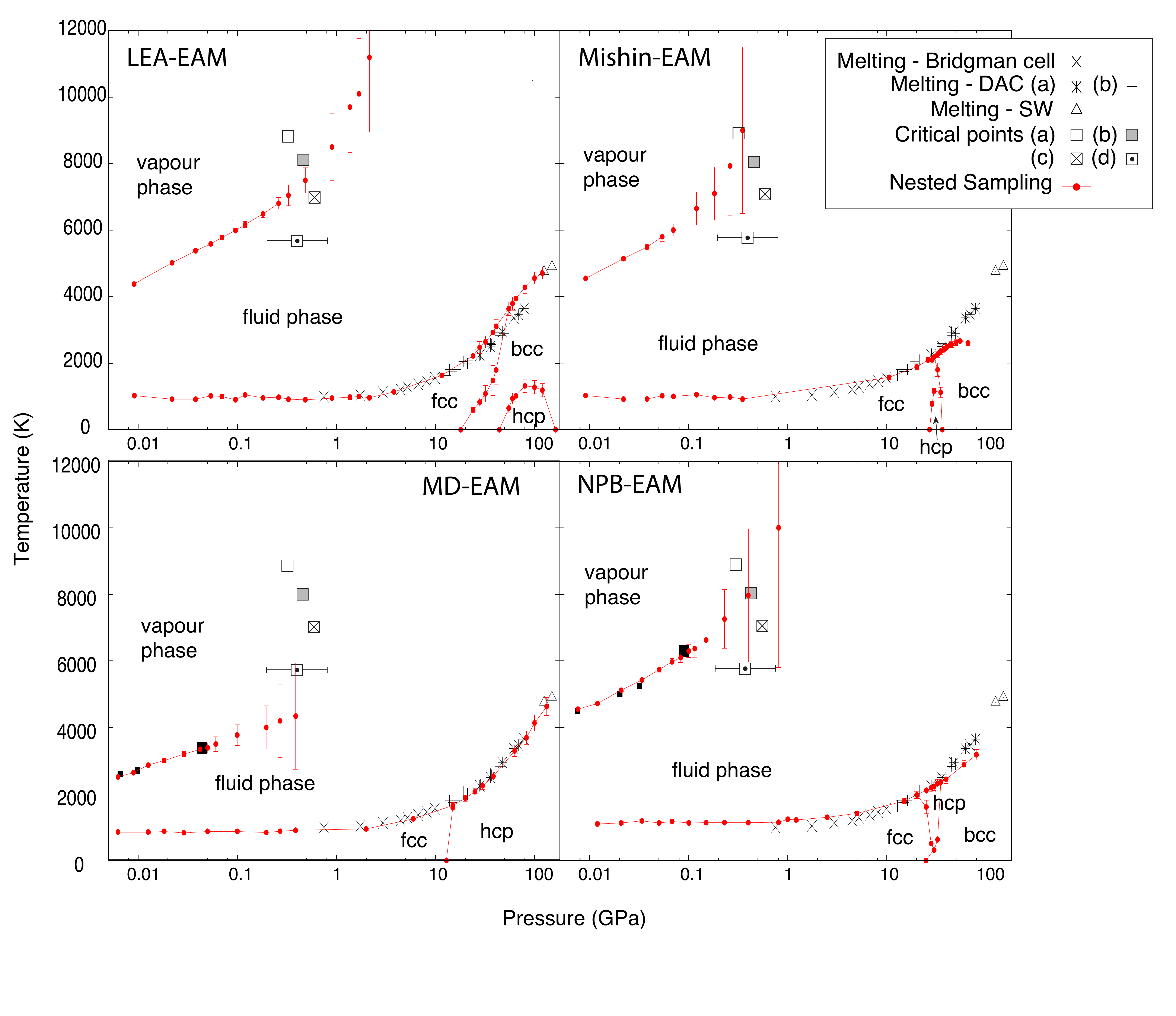}
\vspace{-40pt}
\caption {Phase diagrams corresponding to four 
EAM models of aluminium. Red symbols show the NS results, the error bars are
calculated as the width at half maximum of the peaks on the heat capacity
curves. On the boiling line points  are connected by a solid line up to the critical point. 
(The method we used to estimate the critical point is described in the SM~\cite{SM}.)
Black symbols show experimental melting points measured with Bridgman cells~\cite{Errandonea:2010ct}, with Diamond anvil cells
(DAC (a)~\cite{Boehler:1997vv} and (b)~\cite{Hanstrom:2000tr}) and shock waves (SW)~\cite{bib:Shaner_SWAl}. Different square symbols show estimates
of the critical point from experiments, (a)~\cite{Al_Tc}, (b)~\cite{CT_Fortov}, (c)~\cite{CT_Alder} and (d)~\cite{CT_Renaudin}. For NPB-EAM and MD-EAM large black squares
show the critical point and smaller black squares show the evaporation temperatures, all calculated using Gibbs ensemble Monte Carlo~\cite{Bhatt:2006fw}. At pressures below the critical point, NS parameters $K=800$ and $L=3000$ were used (the total number of energy evaluations was $3\times10^{9}$ for each pressure), while  runs at pressures where solid-solid transitions are present required $K=3200$ and $L=15000$ (total number of energy evaluations were $4 \times 10^{10}$).
}
\label{fig:Al_phase_diagram}
\end{figure*}

In this section we apply the new algorithm to several empirical models of aluminium in order to demonstrate the capability of nested sampling to find solid-solid phase transitions without any prior knowledge of the crystal structures or even the existence of multiple stable phases. Furthermore, although the particular off-the-shelf models we use here do not reproduce the experimentally determined phase diagram of the material everywhere, the fact that nested sampling allows a direct calculation of the entire phase diagram means that in the future one could automate the optimisation of potentials to match the experimental phase diagram. 

 As one of the most commonly used metals, the thermodynamic properties of
aluminium have been extensively studied. The melting line of aluminium has
been measured up to 125 GPa~\cite{Errandonea:2010ct,Boehler:1997vv,Hanstrom:2000tr,bib:Shaner_SWAl},
with good agreement between the different experimental techniques. Theoretical
calculations have also been performed using embedded-atom type potentials~\cite{Foiles_EAM,Voter_EAM,OhJoh_EAM,Mei:1992uc,Morris:1994uz,Ercol_EAM,Liu:2004js,Mishin_EAM}
and  \emph{ab initio} methods~\cite{deWijs:1998vm,Vocadlo:2002ih,Alfe:2004iq},
the latter providing melting temperatures up to 350 GPa~\cite{Bouchet:2009fo}.
At ambient conditions aluminium crystallises in the face-centred-cubic (fcc) structure, but a phase transition to the hexagonal-close-packed (hcp) structure at 217 GPa has been revealed by X-ray diffraction experiments~\cite{Akahama:2006iz} and the body-centred-cubic (bcc) phase has been also produced in laser-induced microexplosions~\cite{Vailionis:2011bs}.
The critical points of most metals are not amenable to conventional experimental study and thus  estimation of  their properties is usually based upon empirical relationships between the critical temperature and other measured thermodynamic properties.
In the case of aluminium these result in predictions in a wide temperature and pressure range~\cite{CT_Renaudin,Al_Tc,CT_Fortov,CT_Alder}.

We chose four widely used models all based on
the embedded-atom method (EAM): (1) the model developed by Liu {\em et al.}~\cite{Liu:2004js} (LEA-EAM), which is an improved version of the original potential of Ercolessi and Adams~\cite{Ercol_EAM}, (2) the model developed by Mishin {\em et al.}~\cite{Mishin_EAM}  using both experimental and \emph{ab initio} data (Mishin-EAM), (3) the EAM of Mei and Davenport~\cite{Mei:1992uc} (MD-EAM) and (4) the recently modified version of the MD-EAM, reparametrised by Jasper {\em et al.} to accurately reproduce the DFT energies for Al clusters and nanoparticles of various sizes  (NPB-EAM)~\cite{NPB_EAM}.

The phase diagrams for all four models based on NS simulations with 64 particles are shown in Figure~\ref{fig:Al_phase_diagram}.
The resulting critical parameters vary over a wide range for the different  models.
Above the critical point we observe the Widom-line, indicated by those points not linked by a solid line. Heat capacity maxima corresponding to the Widom-line become broader away from the critical point, as indicated by the larger error bars. The Widom-line and our method for estimating the critical point are described in the SM~\cite{SM}.

The melting lines are in a good agreement with the available experimental data up to the pressure value $p \approx 25$ GPa. Above that pressure, the melting curves diverge from the experimental results, except for the MD-EAM potential, which reproduces the melting curve remarkably well.

At higher pressures small peaks appear on the heat capacity curves  below the melting temperature for all models indicating  solid-solid phase transitions (see appendix~\ref{app:solid-solid_pt}). 
We post-processed the samples from the NS simulations.
As expected, the fcc structure was found to be stable at low pressures in all four models.
However, the models differ markedly in their predictions at high pressures. 
The only commonality between the predicted high pressure solid phase diagrams is that the maximum predicted stable pressure for the fcc structure is far too low, both in comparison  with experiment and density functional theory~\cite{Akahama:2006iz,Boettger_hcp,Sinko_hcp}.

\subsection{NiTi}\label{subsec:niti}

\begin{figure}[tb]
\begin{center}
\includegraphics[width=8.5cm]{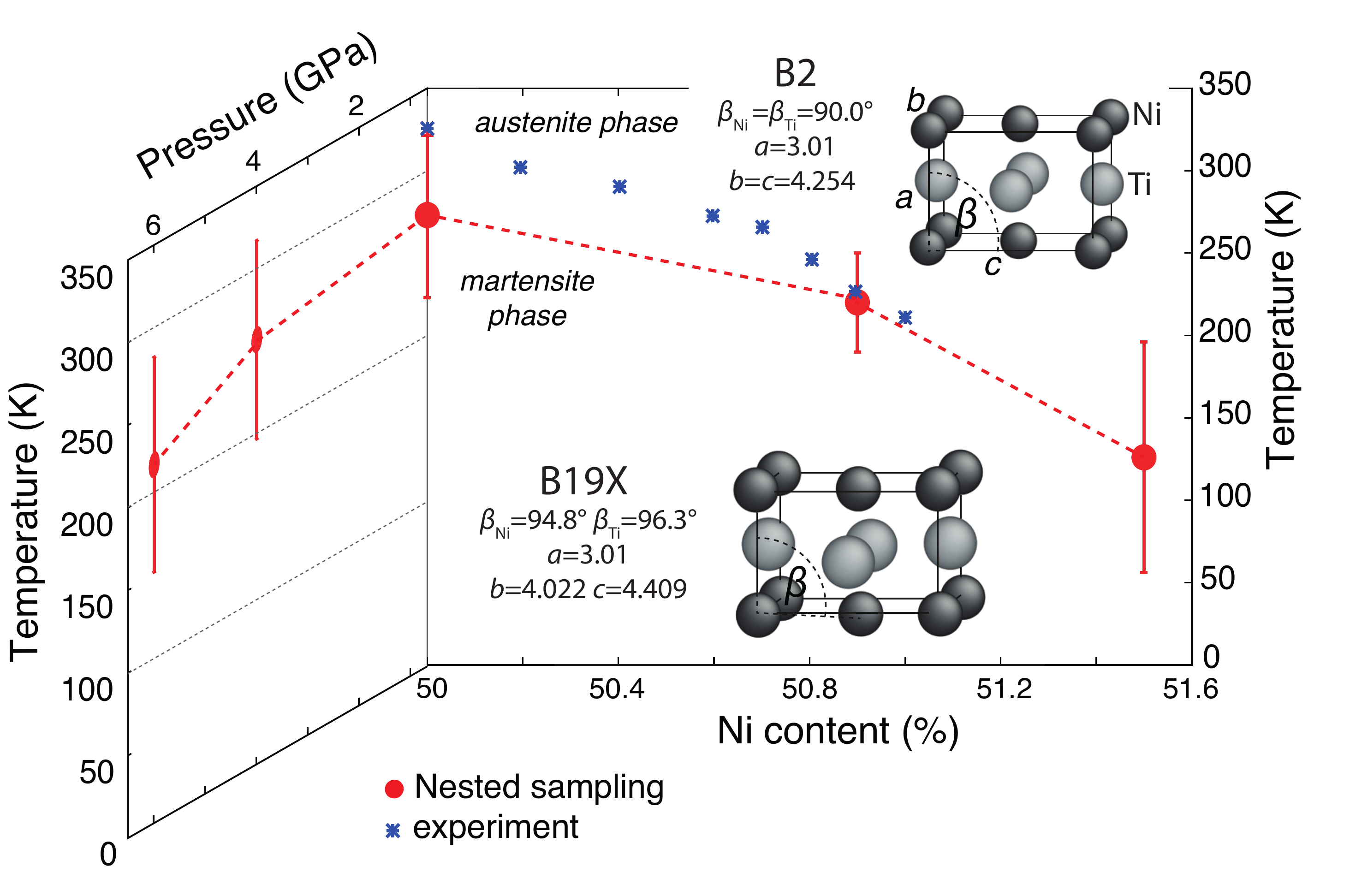}
\end{center}
\vspace{-10pt}
\caption {NiTi martensitic phase transition as a function of Ni content (at 0.66 GPa) and pressure (at 1:1 composition). The simulation cell contained 64 atoms in the cases of the 50\% and 51.6\% Ni compositions and 108 atoms in the case of 50.8\% Ni content. NS parameters were $K=1920$,  $L=10^5$, each data point used $10^{10} $ energy evaluations, and Ni--Ti swap moves were also included in the MC. Experimental results are taken from~\cite{NiTi_exp}.}
\label{fig:NiTi}
\end{figure}

Finally, in order to demonstrate that NS is applicable to more complex problems, we show results for a material of current scientific interest, the NiTi shape memory alloy~\cite{NiTi_Ji, NiTi_discover}. 
The shape memory effect relies on the structural phase transition from the high temperature austenitic phase  to the low temperature martensitic phase~\cite{NiTi_transition}.
Studying this transition is particularly challenging with traditional free energy methods because the austenitic phase does not correspond to a local minimum of the potential energy surface. 
Figure~\ref{fig:NiTi} shows the pressure-temperature-composition phase diagram corresponding to a recent EAM model~\cite{NiTi_EAM, NiTi_strct_trans} as computed with NS. 
The phase transition temperatures are within 50 K of the experimental values and  reproduce the trend with compositional change. 
We predict a decreasing transition temperature with increasing pressure. 
It is notable that this potential successfully reproduces the 
martensitic transition temperature, despite the fact that the minimum enthalpy structure for the potential is different to the structure observed both experimentally and in DFT: here the lowest enthalpy structure (which we label B19X) is orthorhombic (see the SM~\cite{SM} for a description of the low enthalpy structures we identified).
Thus it appears that the austenite-martensite transition temperature is not sensitive to the detailed geometry and ordering of the lowest enthalpy structures. 
Such empirical potentials can therefore be useful tools for studying this transition in the future.

\section{Conclusion and outlook}\label{sec:conc_outlook}

In summary, we have extended the nested sampling algorithm to allow simulations 
using fully flexible periodic boundary conditions at fixed pressure
and demonstrated how it can be used to determine  pressure-temperature-composition phase diagrams. In contrast to existing methods for comparing specific phases, NS explores the entire configuration space without requiring any prior knowledge about the structures of different solid phases with the only necessary input
being the composition and the desired pressure and temperature ranges. This makes it the method of choice for exploring the pressure-temperature-composition space, which is the next unexplored realm naturally following much recent work in crystal structure exploration at zero temperature. Since the algorithm is run independently for different pressures and compositions, and also has excellent parallel scaling up to a number of processors equal to the number of simultaneous samples, it might even be possible to run it on ab initio models on exascale computers. Furthermore, we suggest NS is eminently suitable for validating materials models, and in the future could even play a role in the automatic optimisation of empirical models. 

\begin{acknowledgements}
RJNB acknowledges support from the EPSRC. LBP acknowledges support from St. Catharine's College, Cambridge and to the Royal Society. APB acknowledges support from Magdalene College, Cambridge, the Leverhulme Trust and the Isaac Newton Trust. GC acknowledges EPSRC grant EP/J010847/1. Computer time was provided via the UKCP consortium funded by EPSRC grant ref. EP/K013564/1 and via the Darwin supercomputer in the University of Cambridge High Performance Computing Service funded under EPSRC grant EP/J017639/1. 
Data from this publication can be found at \texttt{https://www.repository.cam.ac.uk/handle/\\1810/255091}.
\end{acknowledgements}

\begin{appendices}

\section{Ideal gas contribution to the partition function}\label{app:ig_pf}

In this appendix we show that the ideal gas contribution to the partition function~\eqref{eq:int_split} asymptotically approaches zero for any positive minimum cell depth $\mch$, in the limit $ k_{\mathrm{B}}T / p V_0 \rightarrow 0 $.

The ideal gas contribution to the partition function~\eqref{eq:int_split} is
\begin{equation}
\int_{\ch\left(\mathbf{h}_0\right) > \mch} \!\!\!\!\!\!\!\!\! d\mathbf{h}_{0}  \delta\left(\det{\mathbf{h}_{0}}-1\right) \int_{V_0}^{\infty} \!\!d V V^N  \int_{\left(0,1\right)^{3N}} \! d\mathbf{s} \, e^{-\beta pV } 
\end{equation}
We begin by noting that the exponential term does not depend on $E(\mathbf{s},\mathbf{h}_0,V)$, and therefore $\int_{(0,1)^{3N}} d \mathbf{s} = 1$.
Thus we have
\begin{equation}
\begin{aligned}
\int_{\ch\left(\mathbf{h}_0\right) > \mch} \!\!\!\!\!\!\!\!\! d\mathbf{h}_{0}  \delta\left(\det{\mathbf{h}_{0}}-1\right) \int_{V_0}^{\infty} d V V^N \int_{(0,1)^{3N}}  \mathrm {d} \mathbf{s} e^{-\beta pV}  \\
= \int_{\ch\left(\mathbf{h}_0\right) > \mch} \!\!\!\!\!\!\!\!\! d\mathbf{h}_{0}  \delta\left(\det{\mathbf{h}_{0}}-1\right) \int_{V_0}^{\infty} d V  V^N e^{-\beta pV}  . 
\end{aligned}
\end{equation}

The  integral over volume $V$ evaluates to
\begin{equation} 
\int_{V_0}^{\infty} d V  V^N e^{-\beta pV} = \frac {1} {(\beta p)^{N+1}} \Gamma(N+1,\beta p V_0)
\end{equation}
where $\Gamma(N+1,\beta p V_0)$ is the upper incomplete gamma function.

Finally 
we
define the function $A\left(\mch\right)$ to be equal to the integral over $\mathbf{h}_{0}$
\begin{equation}
A\left(\mch\right) = \int_{\ch\left(\mathbf{h}_0\right) > \mch} \!\!\!\!\!\!\!\!\! d\mathbf{h}_{0}  \delta\left(\det{\mathbf{h}_{0}}-1\right).
\end{equation}
The function $A\left(\mch\right)$ is finite for any positive value of $\mch$, $A\left( 1	\right)=0$ and $A\left(\mch\right)$ diverges in the limit $\mch \rightarrow 0$.
In the orthorhombic case, where all angles of the simulation cell are equal to $\frac{\pi}{2}$,
$A\left(\mch\right) =  \frac{9}{2}\left(  \log\mch   \right)^2$, with $A=1$ at $\mch \approx 0.62$.
However at any positive value of $\mch$ the contribution to the partition function~\eqref{eq:int_split} due to volumes greater than $V_0$ goes to zero in the limit $k_{\mathrm{B}}T / p V_0 \rightarrow 0$ because $\Gamma(N+1,\beta p V_0) \rightarrow 0$ in the same limit.

\section{Identifying solid-solid phase transitions}\label{app:solid-solid_pt}

The locations of phase transitions are determined solely by looking at the peaks in the heat capacity.
Next, we inspect the system at temperatures either side of the phase transition.
Specific phases can be identified in the following way.
If no appropriate order parameter is to hand, then
one picks a number of random configurations from the output of nested sampling, chosen according to their thermal weights $ \Delta\chi_i e^{-\beta H_i} $, and inspects them by eye.
If an appropriate order parameter {\em is} known, one can compute the free energy landscape for that order parameter. 
Here one proceeds by binning  the weights $ \Delta\chi_i e^{-\beta H_i} $ of all configurations, according to the order parameter, to create a partial sum $\Delta_j=\sum{\Delta\chi_i e^{-\beta H_i}}$ for each bin $j$.
The free energy for each bin can then be computed as $F_j = -\frac{1}{\beta}\left[\log(\Delta_j) + \log\left(\frac{\beta p}{N!}\right) + \frac{3N}{2}\log\left( \frac{2\pi m}{\beta h^2} \right) \right]$.  
In fact, simply calculating the expected enthalpy at the phase transition, and then examining the order parameter values for output configurations around that enthalpy is often sufficient to identify the crystal structures.

\begin{figure}[h]
\begin{center}
\includegraphics[width=9cm]{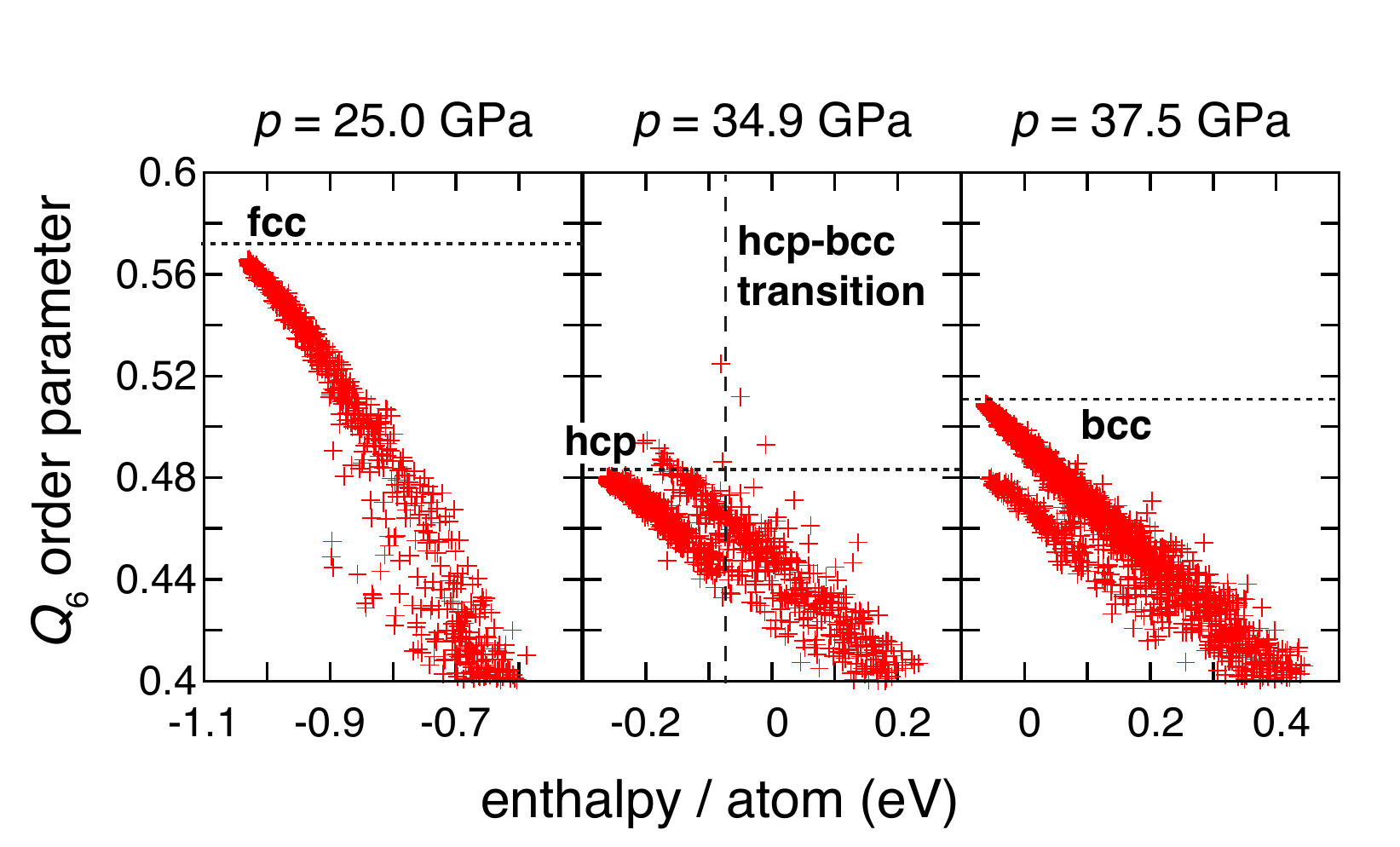}
\end{center}
\caption {Identification of solid phases from inspection of an order parameter: Mishin-EAM potential for aluminium. 
These plots compare the enthalpies and $Q_6$ bond order parameter values for nested sampling output configurations at three different pressures.
Nested sampling proceeds towards lower enthalpies, thus from right to left in each plot.
Horizontal dotted lines show the bond order parameters for the perfect fcc, bcc and hcp phases, and the vertical dashed line shows the expected enthalpy at the solid-solid phase transition, which was located by inspecting the heat capacity and observing a peak.
These results are discussed in the text.
}
\label{fig:Al_NSQ6}
\end{figure}

An example of the latter approach is shown in Figure~\ref{fig:Al_NSQ6} for the Mishin-EAM potential, which compares the enthalpies and $Q_6$ bond order parameter values for nested sampling output configurations at three different pressures.
At $p=25.0$ GPa no phase transition occurs, and only fcc configurations are present. 
At $p=34.9$ GPa a first order phase transition occurs at the average enthalpy marked by the vertical dashed line. 
At that enthalpy there is a clear transition between two basins, from a first basin 
that corresponds to the bcc structure, to a second 
that corresponds to the hcp structure.
Finally, at $p=37.5$ GPa no phase transition occurs and so there is no peak in the heat capacity.
At this pressure the bcc structure is stable at all temperatures below the melting point.
Nevertheless, the hcp structure is clearly visible as a metastable structure.

\restoreapp
\end{appendices}

\onecolumngrid
\clearpage

\begin{center}
\textbf{\large Determining pressure-temperature phase diagrams of materials: Supplemental Material }
\end{center}

\setcounter{equation}{0}
\setcounter{figure}{0}
\setcounter{section}{0}
\setcounter{subsection}{0}
\setcounter{subsubsection}{0}
\setcounter{table}{0}
\setcounter{page}{1}
\makeatletter
\renewcommand{\theequation}{S\arabic{equation}}
\renewcommand{\thefigure}{S\arabic{figure}}
\renewcommand{\thetable}{S\arabic{table}}
\renewcommand{\bibnumfmt}[1]{[S#1]}
\renewcommand{\citenumfont}[1]{S#1}

\section{Detailed Monte Carlo Scheme}\label{app:mc_scheme}

To understand the Monte Carlo scheme for NS in the fixed pressure ensemble with a fully flexible cell, it is helpful to first consider the Monte Carlo scheme for the fixed volume ensemble,  with a fixed cell.
In that case the partition function is given by
\begin{equation}
Z(N,V,\beta) = \frac{1}{N!}\left(\frac{2\pi m}{\beta h^2}\right)^{3N/2} V^N \int d\mathbf{s} \, e^{-\beta E\left( \mathbf{s}, V \right)} . \label{eq:Z_NVT_2} 
\end{equation}
As described in previous papers~\cite{bib:our_NS_paper_sm,Burkoff1_sm,bib:Frenkel_NS_sm}, each iteration begins with $K$ samples distributed uniformly in $\mathbf{s}$, with $E<E_{\mathrm{limit}}$.
That is to say, those $K$ samples are drawn from the distribution
\begin{equation} \label{eq:configs_unif_below_cutoff_NVT}
\mathrm{Prob}\left(\mathbf{s}\vert E_{\mathrm{limit}} \right) = 
\begin{cases}
\frac{1}{\chi\left( E_{\mathrm{limit}} \right) }, & E\left(\mathbf{s}, V \right)  < E_{\mathrm{limit}}   \\
0 , & \text{Elsewhere}.
\end{cases}
\end{equation}

For the fixed pressure ensemble,
we must explore volume and cell matrices $\mathbf{h}_0$, as well as the fractional coordinates $\mathbf{s}$.
In particular,  we calculate the integral (9) with NS.
\begin{equation}
 \Delta_{\mathrm{NS}} (N,p,\beta,V_0,\mch)  = \int_{\ch\left(\mathbf{h}_0\right) > \mch} \!\!\!\!\!\!\!\!\! d\mathbf{h}_{0}  \delta\left(\det{\mathbf{h}_{0}}-1\right) \times  \int_{0}^{V_0} \!\!d V V^N \int_{\left(0,1\right)^{3N}}\! d\mathbf{s} \, e^{-\beta\left[  E\left( \mathbf{s},\mathbf{h}_{0},V \right) + pV \right]} 
\tag{ 9, revisited}
\end{equation}
In (9) the volume has a weight $V^N$ with $V<V_0$, and $\mathbf{h}_0$ is constrained to a surface of determinant $1$, with $\ch\left(\mathbf{h}_0\right) > \mch$.
Therefore we seek to generate new samples according to the distribution
\begin{equation} \label{eq:configs_unif_below_cutoff_NPT}
\begin{aligned}
& \mathrm{Prob}\left(\mathbf{s}, \mathbf{h}_{0}, V \vert H_{\mathrm{limit}} \right)  \\ & = 
\begin{cases}
\frac{V^N \delta\left(\det{\mathbf{h}_{0}}-1\right)}{\chi\left( H_{\mathrm{limit}} \right) }, & H\left(\mathbf{s}, \mathbf{h}_{0},  V \right)  < H_{\mathrm{limit}} , \: V < V_0 , \\ & \ch\left(\mathbf{h}_0\right) > \mch  \\ 
0 , & \text{Elsewhere}.
\end{cases}
\end{aligned}
\end{equation}

As described in the paper, this can be achieved by Markov chain Monte Carlo (MCMC) sampling.
MCMC walks were random sequences of the following Monte Carlo (MC) moves, which were chosen according to the probabilities (ratios) given in the paper, unless specifically stated in this Supplemental Material.

\begin{description}
\item[Atom moves] Single atom MC moves are performed as follows.
\begin{enumerate}
\item Select a single atom at random.
This atom has fractional coordinates $\mathbf{s}_i$
\item Displace that atom's fractional coordinates by a random vector
\begin{equation}
\mathbf{s}_i \rightarrow \mathbf{s}_i + L_s \mathbf{\Delta_s}
\end{equation}
Here $\mathbf{\Delta_s}$ is a 3-tuple with elements drawn independently from a uniform distribution on the interval $[-1,1]$, and $L_s$ is a constant that controls the step size.
The size of $L_s$ was updated periodically during the NS calculation, as described in the next subsection, ``Updating MC step lengths''.
\item Calculate the enthalpy $H_{\mathrm{trial}} = E\left( \mathbf{s},\mathbf{h}_{0},V \right) + pV $ for the system after this atom displacement.
\item Accept the displacement if $H_{\mathrm{trial}}<H_{\mathrm{limit}}$.
Otherwise, the old configuration is kept.
\end{enumerate}

\item[Volume moves]
Volume MC steps are as follows.
\begin{enumerate}
\item Propose a volume move from $V$ to $V'$ where
\begin{equation}
V' = V + L_V \Delta_V
\end{equation}
Here $\Delta_V$ is a random number drawn from the uniform distribution on the interval  $[-1,1]$, and $L_V$ is a constant that controls the step size.
The size of $L_V$ was updated periodically during the NS calculation, as described in the next subsection, ``Updating MC step lengths''.
\item The displacement is accepted with probability $\min\left(1, \left[\frac{V'}{V}\right]^N\right)$.
If the displacement is accepted, continue to step 3.
Otherwise, the old configuration is kept and the move ends.
\item If the displacement was accepted in step 2, calculate the enthalpy $H_{\mathrm{trial}} = E\left( \mathbf{s},\mathbf{h}_{0},V' \right) + pV' $ for the system after this volume displacement.
\item If $H_{\mathrm{trial}}<H_{\mathrm{limit}}$ the displacement is accepted. 
Otherwise, the old configuration is kept.
\end{enumerate}

\item[Lattice shear moves]

The unit simulation cell $\mathbf{h}_0$ is sheared, subject to constraints on both the cell depth and the enthalpy.
The algorithm is given in Algorithm \ref{alg:h0_shear_mc}.
The size of shear steps is controlled by the parameter \texttt{step\_l\_sh}.
This parameter was updated periodically during the NS calculation, as described in the next subsection, ``Updating MC step lengths''.

\item[Lattice stretch moves] 

The unit simulation cell $\mathbf{h}_0$ is stretched at fixed unit volume, subject to constraints on both the cell depth and the enthalpy.
The algorithm is given in Algorithm \ref{alg:h0_stretch_mc}.
The size of stretch steps is controlled by the  parameter \texttt{step\_l\_st}.
This parameter was updated periodically during the NS calculation, as described in the next subsection, ``Updating MC step lengths''.

\item[(Binary) atom swaps]

For the NiTi alloy, we included binary (Ni--Ti) atom swaps as an additional MC proposal.
The algorithm for this MC proposal is as follows:

\begin{enumerate}
\item Choose a random Ni atom, and a random Ti atom.
\item Swap the fractional coordinates of those two atoms.
\item Calculate the enthalpy $H_{\mathrm{trial}} = E\left( \mathbf{s},\mathbf{h}_{0},V \right) + pV $ for the system after the swap.
\item Accept the swap if $H_{\mathrm{trial}}<H_{\mathrm{limit}}$.
Otherwise, the old configuration is kept.
\end{enumerate}

\end{description}

\begin{alg}[caption={[Lattice shear MC step subject to minimum cell depth criterion] Lattice shear MC step. Samples the surface $\det(\mathbf{h}_0) = 1$ subject to the enthalpy and minimum cell depth constraints. $\mathtt{ranf()}$ is a random number uniform in [0,1].},label=alg:h0_shear_mc][h]
subroutine lattice_shear_mc
! lattice shear MC step
! The input unit cell matrix is h0_in(3,3)

h0_trial = h0_in              ! copy h0_in into working array

do repeat = 1, 3              ! repeat three times

  h0_save = h0_trial          ! save $\mathbf{h}_0$ in case of rejection

  chng_vec = ranf()*3 + 1     ! randomly selected lattice
			      ! vector. integer: 1, 2, or 3

  basis = orthonormal_basis(h0_in,chng_vec)
  ! basis(3,2) is a pair of orthonormal vectors in the plane 
  ! defined by the two edge vectors in h0_in other than chng_vec 

  pair: do
    x = 2.0*(ranf() - 0.5)    ! generate (x,y) pair from 
    y = 2.0*(ranf() - 0.5)    ! inside unit circle 
    if (x**2 + y**2 < 1.0) exit pair
  end do pair
  x = x*step_l_sh             ! x,y updated to uniform pair in
  y = y*step_l_sh             ! a circle of radius step_l_sh

  h0_trial(:,chng_vec) = h0_trial(:,chng_vec) 
			 + x*basis(:,1) + y*basis(:,2)
  ! change to cell vector chng_vec is circularly symmetric 
  ! and in the plane defined by the other two edge vectors

  if ( ch(h0_trial) < mch ) h0_trial = h0_save
  ! trial rejected due to minimum cell depth criterion.
  ! cell depth of proposed cell is smaller than minimum
  ! allowed value: $ \ch\left(\mathbf{h}_0\right) < \mch $

end do

H_trial = H(h0_trial)     ! H = configurational enthalpy
if (H_trial <= H_lim) then
  h0_out = h0_trial       ! accept trial
else
  h0_out = h0_in          ! trial rejected
			  ! $\mathbf{h}_0$ unchanged: a null step
end if

end subroutine
\end{alg}

\begin{alg}[caption={[aasdasd] Lattice stretch MC step. Samples the surface $\det(\mathbf{h}_0) = 1$ subject to the enthalpy and minimum cell depth constraints. $\mathtt{ranf()}$ is a random number uniform in [0,1].},label=alg:h0_stretch_mc][h!]
subroutine lattice_stretch_mc
! lattice stretch MC step
! The input unit cell matrix is h0_in(3,3)

h0_trial = h0_in              ! copy h0_in into working array

do repeat = 1, 3              ! repeat three times

  h0_save = h0_trial          ! save $\mathbf{h}_0$ in case of rejection

  vec1 = ranf()*3 + 1         ! (vec1,vec2) is a random
  vec2 = ranf()*2 + 1         ! pair of different lattice
  if (vec1 == vec2) then      ! vectors
    vec2 = 3
  end if

  ! u is a uniform random variable from 
  ! [-step_l_st,step_l_st]
  u = 2.0*(ranf() - 0.5)*step_l_st

  ! multiply edge vectors vec1, vec2 by exp(u), exp(-u)
  h0_trial = h0_in
  h0_trial(:,vec1) = h0_trial(:,vec1)*exp(u)
  h0_trial(:,vec2) = h0_trial(:,vec1)*exp(-u)

  if ( ch(h0_trial) < mch ) h0_trial = h0_save
  ! trial rejected due to minimum cell depth criterion.
  ! cell depth of proposed cell is smaller than minimum
  ! allowed value: $ \ch\left(\mathbf{h}_0\right) < \mch $

end do

H_trial = H(h0_trial)     ! H = configurational enthalpy
if (H_trial <= H_lim) then
  h0_out = h0_trial       ! accept trial
else
  h0_out = h0_in          ! trial rejected
                          ! $\mathbf{h}_0$ unchanged: a null step
end if

end subroutine
\end{alg}

\subsection{Updating MC step lengths}

The values of the step size parameters, $L_s$, $L_V$, \texttt{step\_l\_sh} and \texttt{step\_l\_st}, were updated every $\frac{K}{2}$ iterations (recall that $K$ is the number of samples employed in the NS calculation).
This corresponds to an expected reduction in the enclosed configuration space volume $\chi$ to a factor of $e^{-\frac{1}{2}}$.
Each step size parameter was updated by performing a short MCMC exploration using only the corresponding type of MC proposal.
The step size parameters were updated to maintain an acceptance rate in the interval $\left(0.2,0.3\right)$, and as close to $0.25$ as possible, for each MC proposal type.

Since the calculation was performed in parallel, when collecting data about the acceptance rate for each MC proposal type, each processor ran MCMC trajectories with separate, random configurations from the current sample set.
In general, this avoids setting step lengths appropriate to unrepresentative regions of the bounded uniform distribution.

\section{Parallelisation scheme}\label{app:Parallelisation}

In order to be able to benefit from supercomputing facilities efficiently, the NS algorithm has to be parallelised. Fortunately, possible parallelisation schemes come quite naturally from the method, and here we describe a scheme for parallelising within each iteration~\cite{bib:Rob_thesis_sm}. While in the serial algorithm, the new clone is decorrelated using a trajectory of walk length $L$ in each iteration, in our scheme parallelised over $N_\mathrm{proc}$ processors, the new clone is decorrelated using a trajectory of walk length $L/N_\mathrm{proc}$, along with  $(N_\mathrm{proc}-1)$ randomly chosen configurations which are also (independently) propagated through $L/N_\mathrm{proc}$ steps. Thus, rather than propagating a single configuration for $L$ steps, we propagate $N_\mathrm{proc}$ configurations for $L/N_\mathrm{proc}$ steps. 

\section{Finite size effect}

\begin{figure}[hbt]
\begin{center}
\includegraphics[width=9cm]{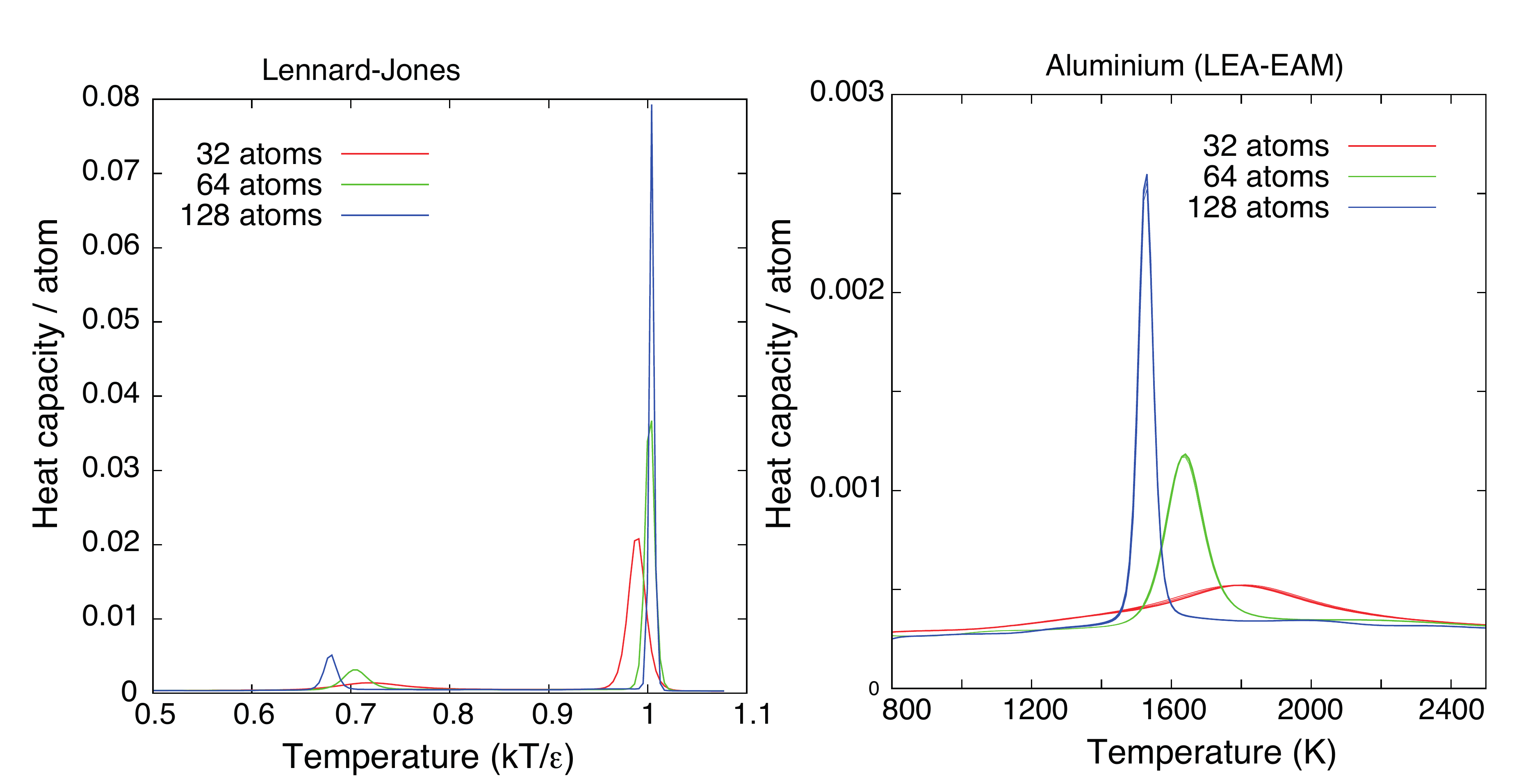}
\end{center}
\vspace{-10pt}
\caption {Magnitude of the finite size effect in case of the Lennard-Jones potential (left panel) run with parameters $K=672$ and $L=15120$, and LEA-EAM aluminium potential (right panel) with run parameters $K=1608$ and $L=10080$, as demonstrated by the heat capacity curves of systems of 32, 64 and 128 particles. }
\label{fig:finite_size}
\end{figure}

In order to estimate the effect of using only 64 particles in our simulations we performed tests with 32 and 128 particles as well for both the periodic Lennard-Jones and the LEA-EAM aluminium potential models. 
The heat capacities of these are compared in Figure~\ref{fig:finite_size}. 
The Lennard-Jones test was performed at $p=0.027 \epsilon/\sigma^3$, below the critical point, hence both the evaporation and melting peaks are present. 
The aluminium test was performed at $p=11.6$~GPa, above the critical point, thus only the melting peak is present. 
The figure shows that increasing the number of atoms, the heat capacity peaks become narrower as expected, 
and the melting and evaporation transitions shift to lower and higher temperatures respectively.
Of course, it would be possible to perform a finite size scaling analysis from the results of multiple Nested Sampling calculations, performed at different system sizes.
However, we felt that the phase diagrams of the various aluminium models
differed from each other to the extent that it was sufficient to give a rough indication of error
based on the full-width-at-half-maximum of the peaks.

\section{Lennard-Jones potentials}

The benchmark comparison to parallel tempering was performed using the truncated and shifted Lennard-Jones potential~\cite{frenkel2001understanding_sm}.
Other Lennard-Jones calculations were performed using the truncated but {\em not} shifted potential with the usual mean-field correction~\cite{frenkel2001understanding_sm} beyond the cut-off radius.
All Lennard-Jones calculations employed a cut-off radius $r_c=3\sigma$.

\section{Aluminium}

\subsection{Simulation details}
The simulation cell contained 64 aluminium atoms. 
The same ratio of the different types of MC moves was used at every pressure.
Sampling is robust with respect to the ratio of MC moves of each type, and for these aluminium calculations the ratio of MC moves (atom:volume:shear:stretch) was (16:1:1:1), rather than (N:10:1:1). 
The minimum allowed cell depth, $\mch$, was set to 0.65. 
Spline parameters for the LEA-EAM and Mishin-EAM potentials are only provided for configurations where no two atoms are closer than a certain, small radius. 
For separations smaller than this radius we made linear extrapolations to the splines. 
The ultra-close atomic proximity region is only relevant to the ultra-high temperature gas phase, and does not contribute to the
thermodynamics of aluminium in the temperature and pressure ranges 
considered in this paper.
Therefore the precise form of this extrapolation does not affect any of the results we show. 

\subsection{Heat capacity curves}
\begin{figure}[h]
\begin{center}
\includegraphics[width=6.8cm]{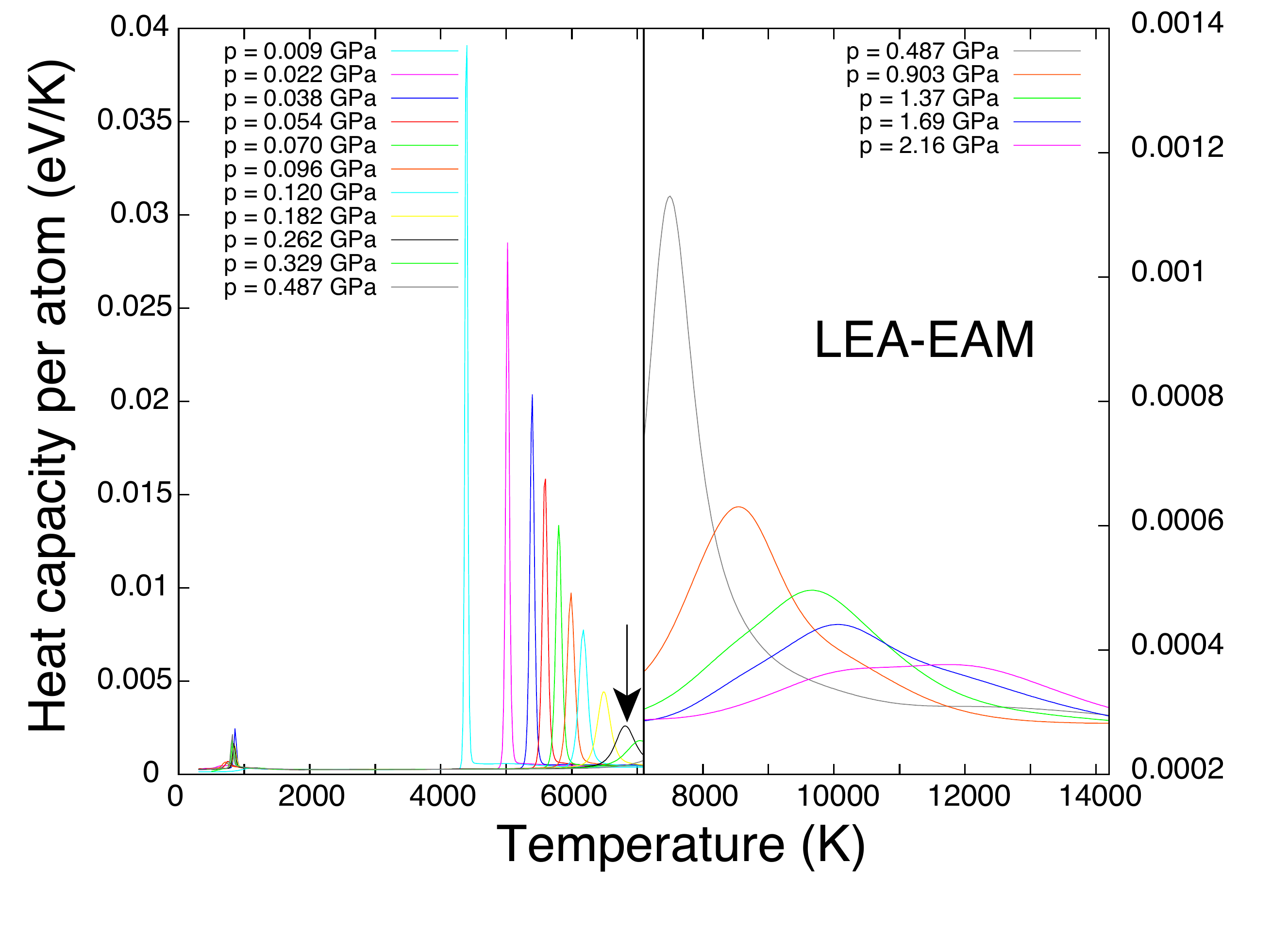}
\includegraphics[width=6.8cm]{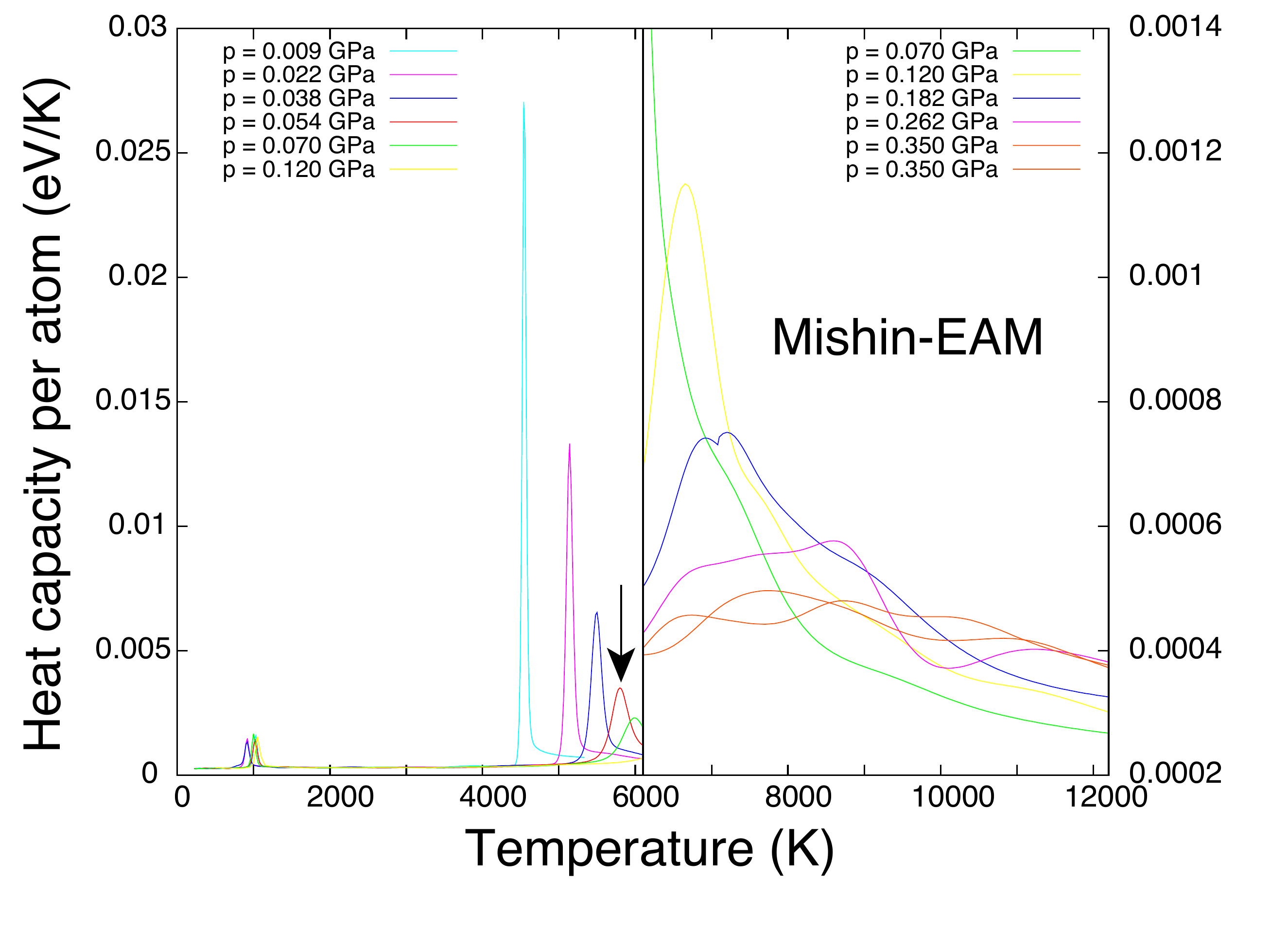}
\includegraphics[width=6.8cm]{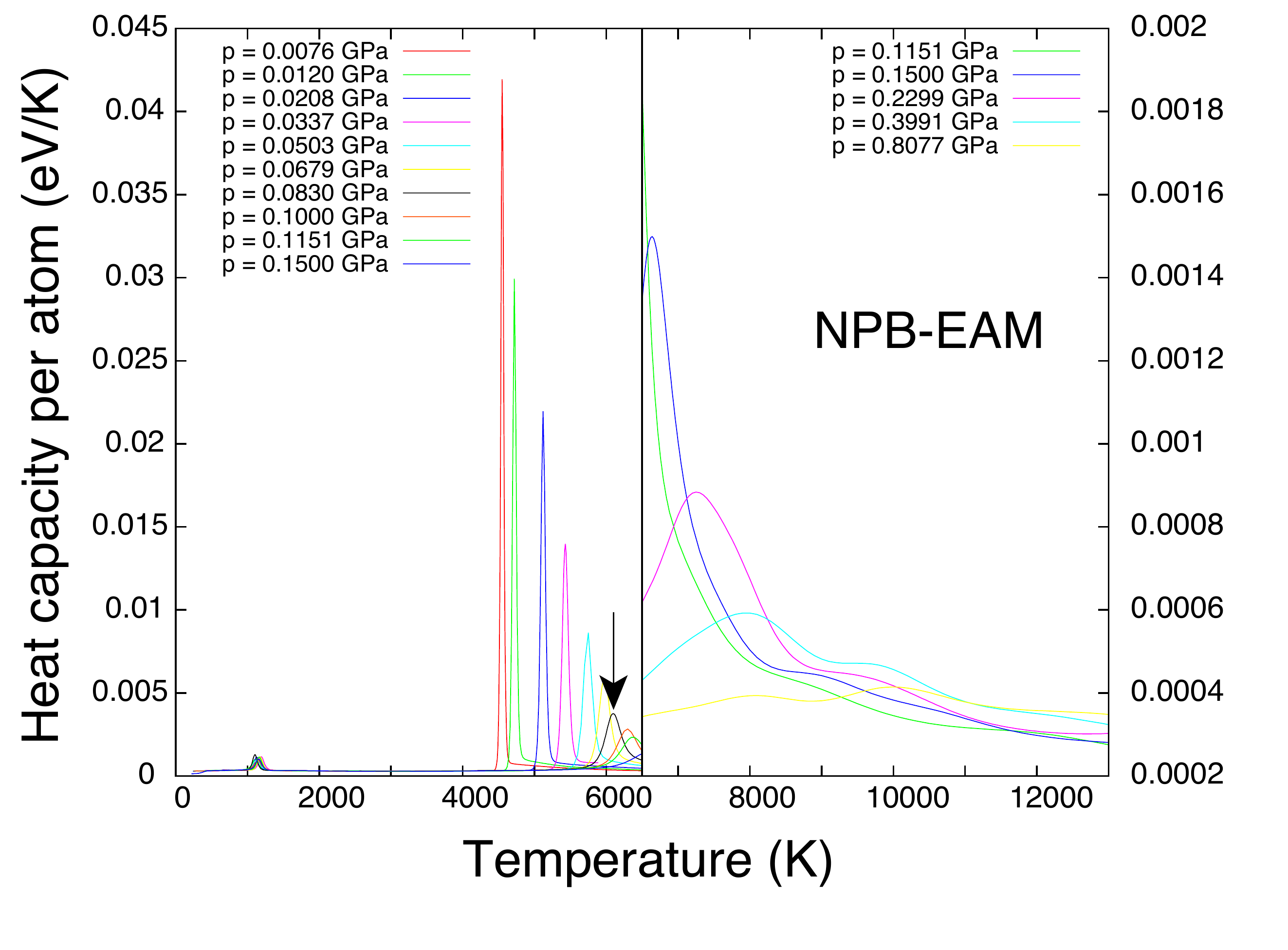}
\includegraphics[width=6.8cm]{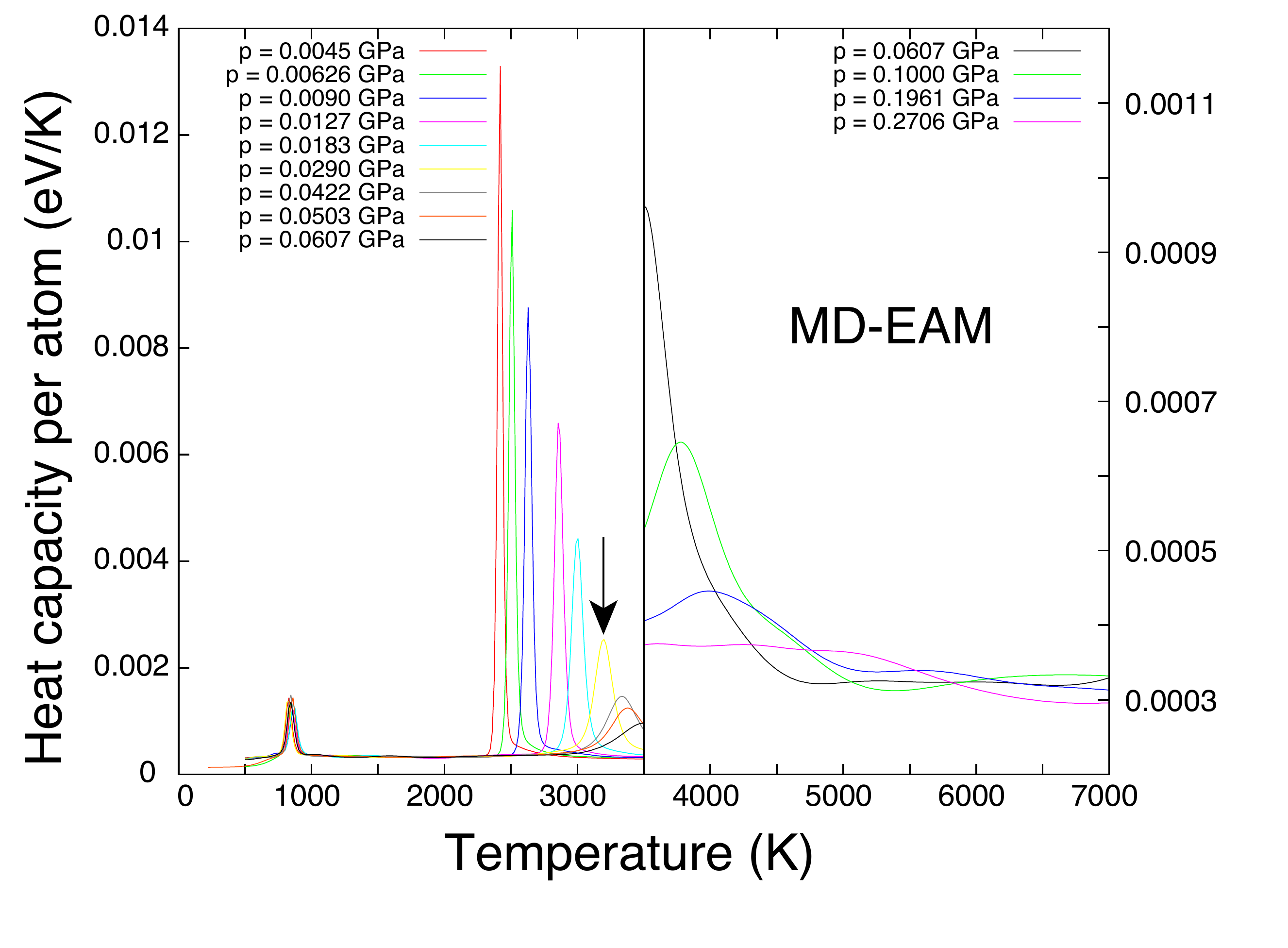}
\end{center}
\vspace{-15pt}
\caption {Aluminium heat capacity curves at lower pressures up to the end of the Widom-line for the different studied potential models. The arrows indicate the peak of the $c_p$ curves closest to the critical point, and in order to show the broadened peaks above that, the curves in the supercritical region are shown on a different scale (shown on the right hand side). The size of the live set and the total number of MC steps during one iteration was $K=804$ and $L=3120$, respectively.}
\label{fig:Al_HC_low}
\end{figure}

\begin{figure}[htb]
\begin{center}
\includegraphics[width=7cm]{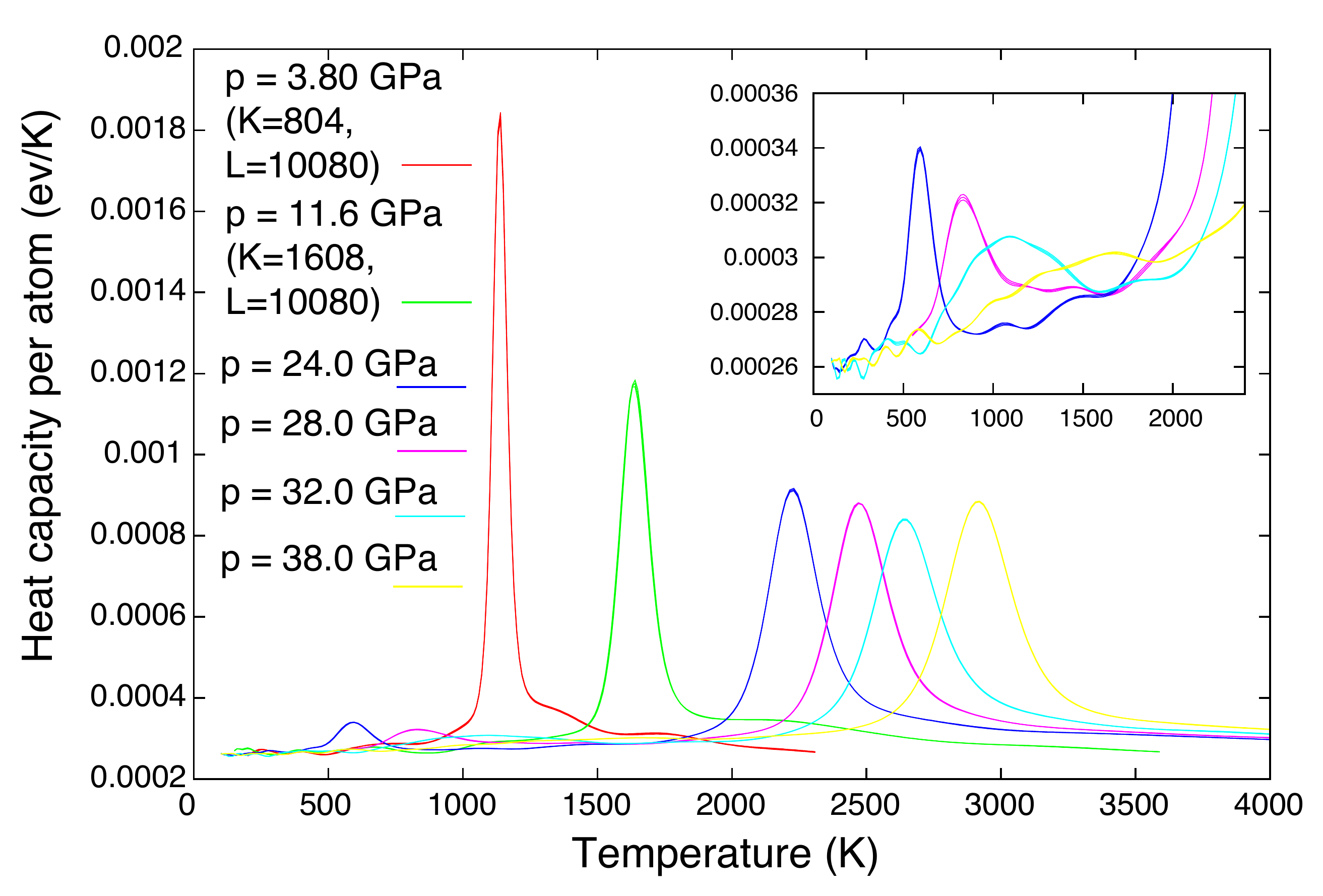}
\includegraphics[width=7cm]{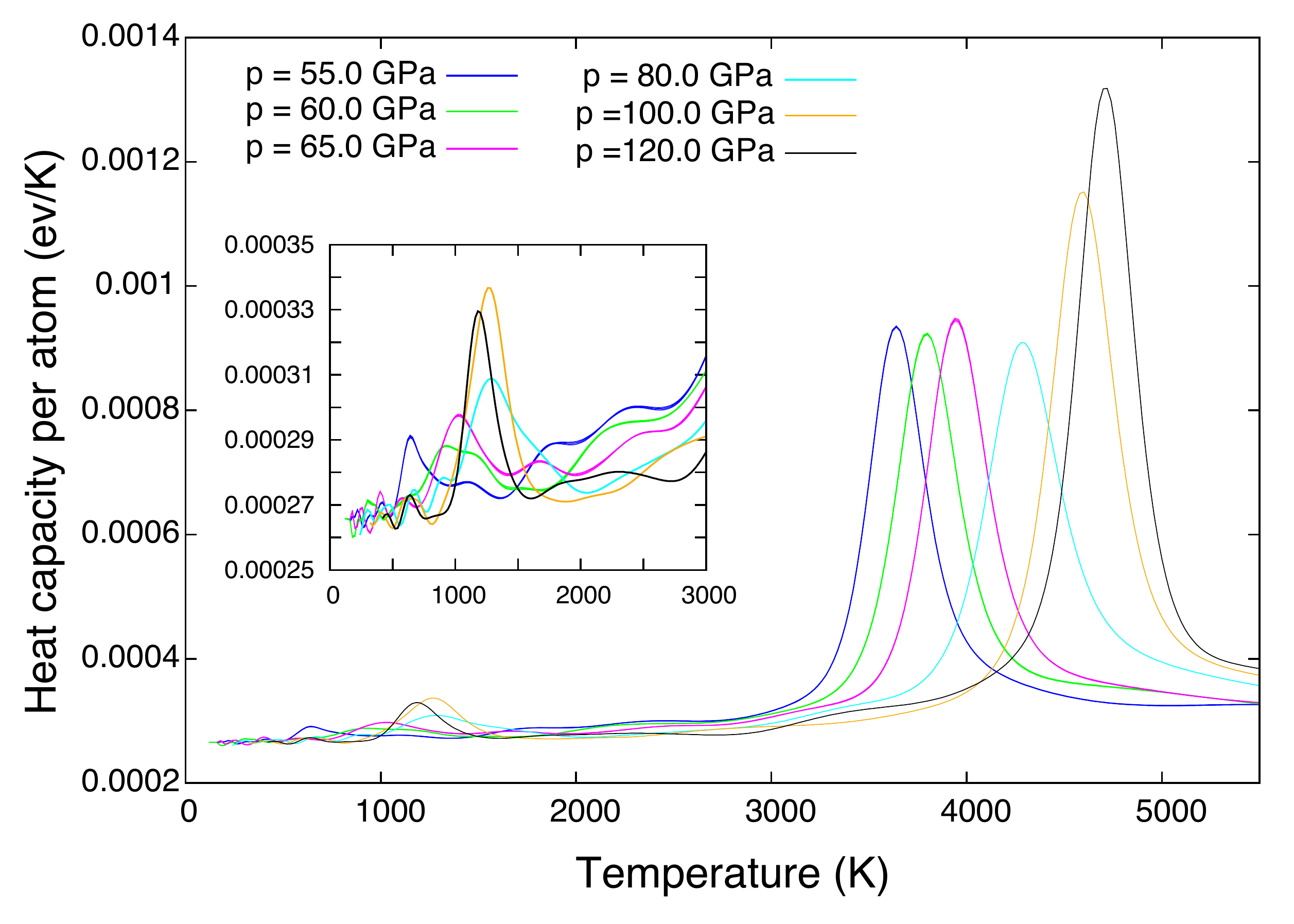}
\end{center}
\vspace{-10pt}
\caption {Aluminium heat capacity curves above the critical point for the LEA-EAM potential. The insets show the solid-solid phase transition peaks in a larger scale. 
The size of the live set was $K=3216$ and the total number of MC steps during one iteration was $L=15120$, or if otherwise, it is indicated in the legend.}
\label{fig:Al_HC_LEA}
\end{figure}

\begin{figure}[hbt]
\begin{center}
\includegraphics[width=7cm]{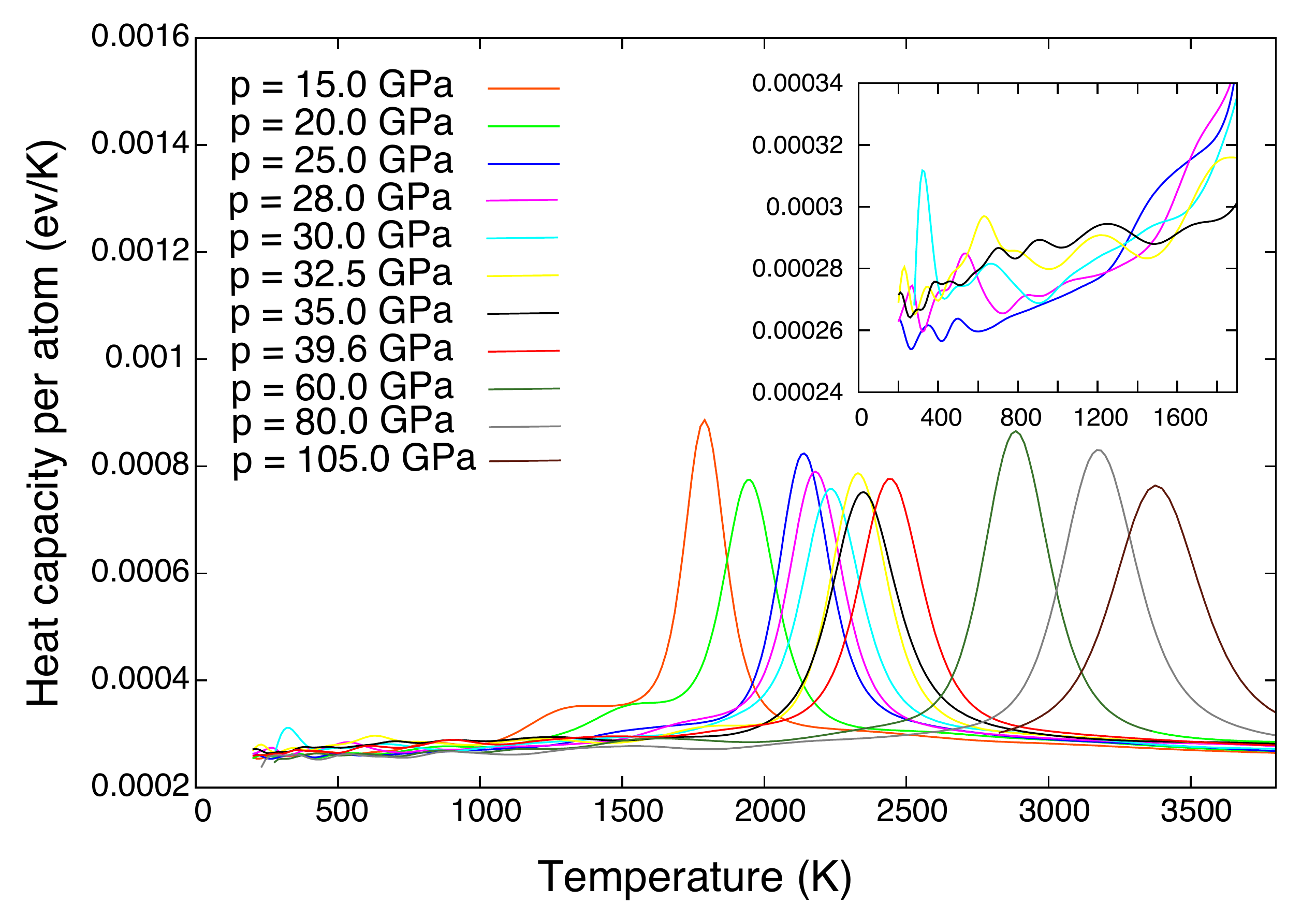}
\end{center}
\vspace{-10pt}
\caption {Aluminium heat capacity curves above the critical point for the NPB-EAM potential. The inset shows the solid-solid phase transition peaks in a larger scale. 
The size of the live set was $K=3216$ and the total number of MC steps during one iteration was $L=15120$.}
\label{fig:Al_HC_NPB}
\end{figure}

\begin{figure}[hbtp]
\begin{center}
\includegraphics[width=7cm]{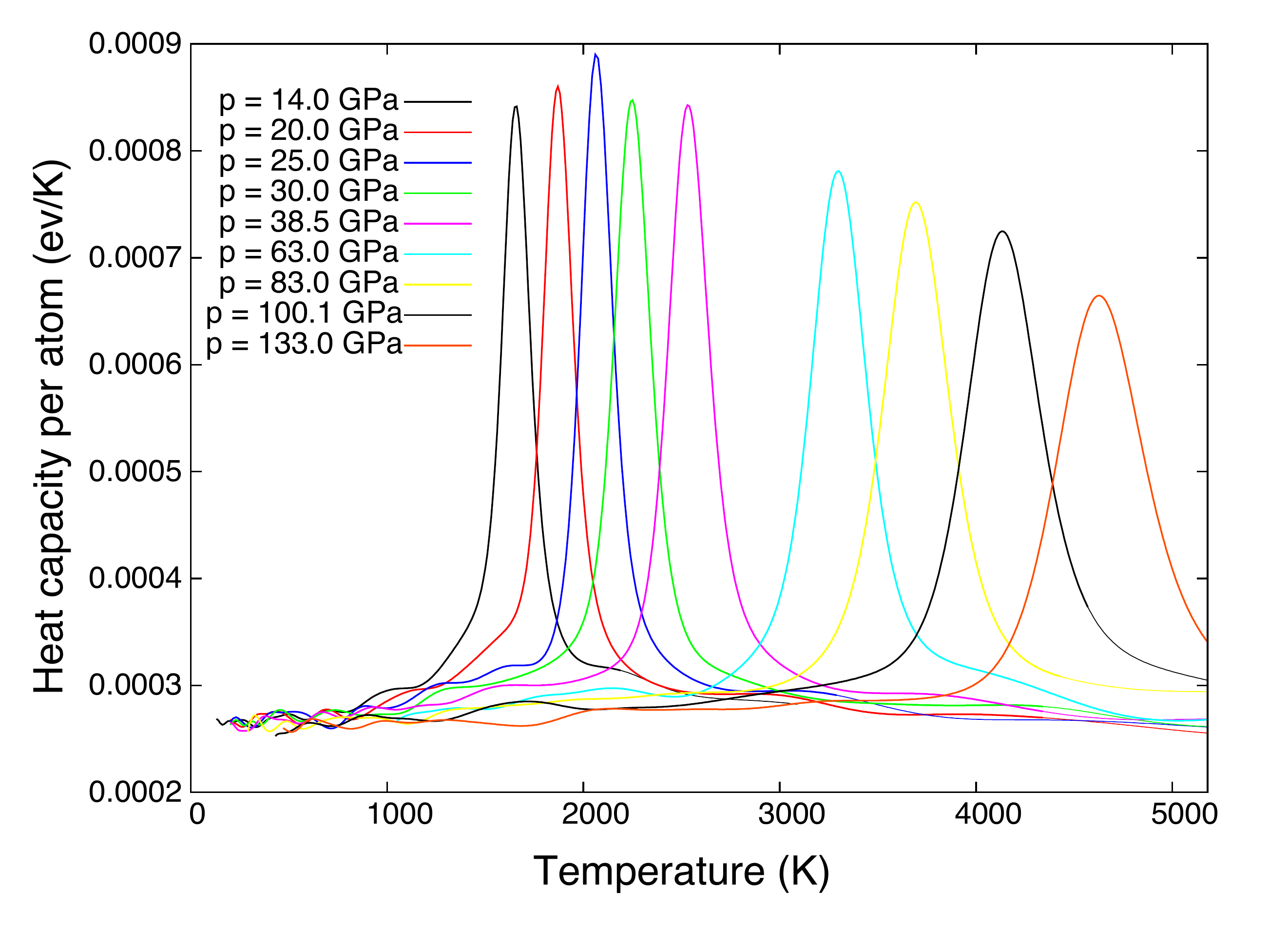}
\end{center}
\vspace{-10pt}
\caption {Aluminium heat capacity curves above the critical point for the MD-EAM potential. 
The size of the live set was $K=3216$ and the total number of MC steps during one iteration was $L=15120$.}
\label{fig:Al_HC_MD}
\end{figure}

Heat capacity curves calculated with Nested Sampling are shown in Figures~\ref{fig:Al_HC_low}, \ref{fig:Al_HC_LEA}, \ref{fig:Al_HC_NPB} and \ref{fig:Al_HC_MD}. 

\subsection{Critical point and the Widom-line}

To locate the critical point, we draw on the results of Bruce and Wilding~\cite{bruce_wilding_scaling_prl_sm}.
Specifically we begin from their analysis of the probability distribution for density for finite systems, in the pressure-temperature region that would contain the critical point in an infinite system. 
For a finite system at and below the critical
point the density distribution for temperatures corresponding to
maxima of the heat capacity appears as a bimodal distribution. 
Above the critical point (along the Widom-line), the distribution transitions quickly to a unimodal distribution. 
In the context of a phase diagram, our location of the critical point appears quite precise, but in fact our sampling of the pressure axis is relatively sparse compared to the width of the scaling region around the critical point, since we are inspecting the system across pressures spanning many orders of magnitude. 
Therefore, between adjacent pressure values the density distribution transitions rapidly between being bimodal and unimodal. 
We chose the lowest pressure for which the distribution is clearly unimodal as an upper bound for the critical pressure. 
The corresponding upper bound for the critical temperature was taken to be the temperature at which the heat capacity, Cp, is its local maximum at that pressure.

According to textbook definitions, beyond the critical point a single fluid phase is defined. However, response functions such as the heat capacity, thermal expansion coefficient and compressibility continue to exhibit maxima into the supercritical region.
Lines of these maxima, which spread out from the critical point are called the Widom-lines.
If one moves away from the critical point, the different Widom-lines rapidly diverge from one another, and the maxima themselves become quickly smeared and disappear, in case of the heat capacity at $T \approx 2.5T_c$ and $p \approx 10 p_c$~\cite{bib:Widom-line_LJ_sm}. The same tendency can be observed in our Nested Sampling results, see Fig.~\ref{fig:Al_HC_low}.

\begin{figure}[htb]
\begin{center}
\includegraphics[width=9cm]{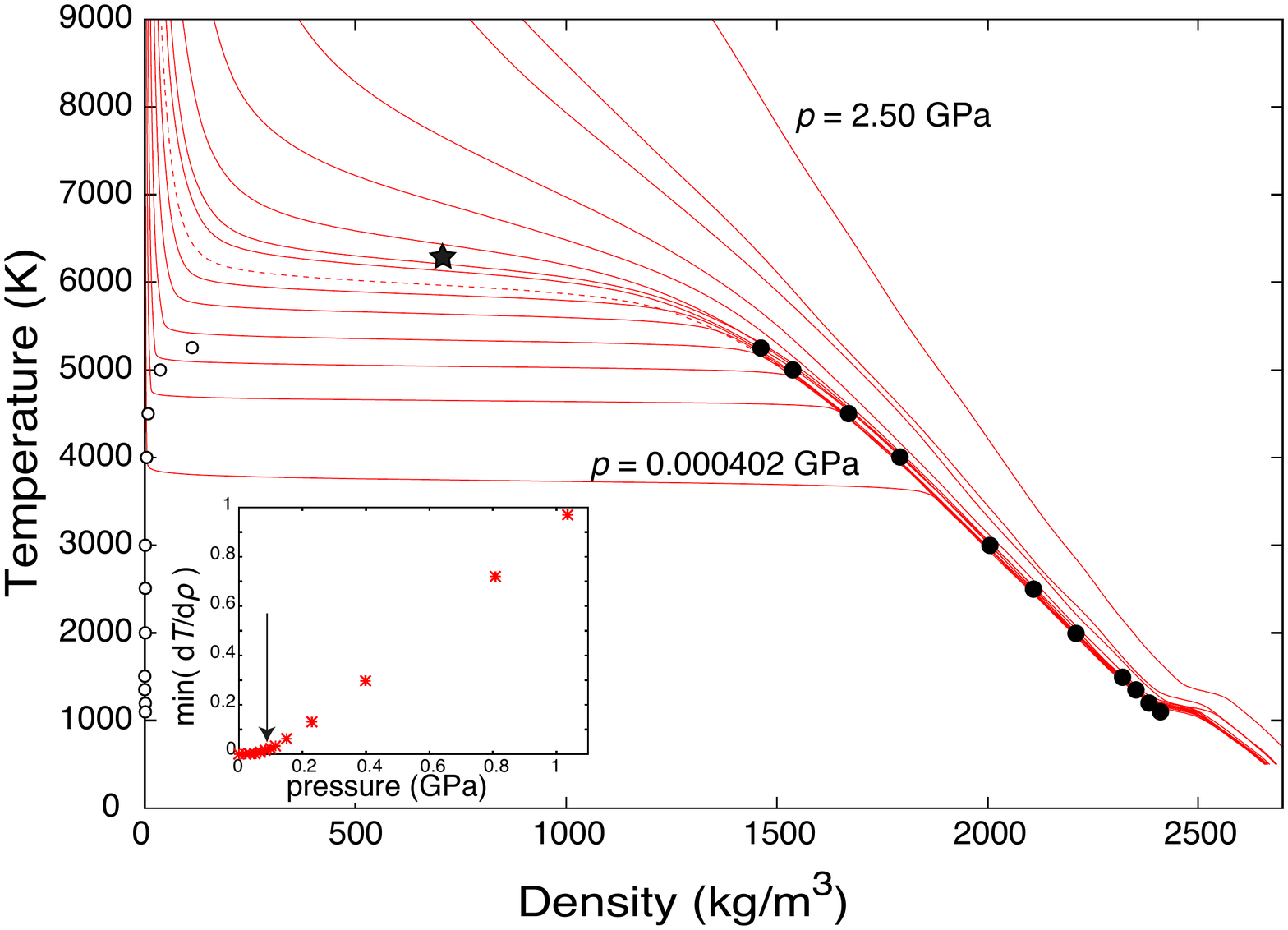}
\end{center}
\vspace{-20pt}
\caption {Density-temperature phase diagram of aluminium with the NPB-EAM potential. Red lines show the Nested Sampling results at different pressures: 0.000402 GPa, 0.0012 GPa, 0.0208 GPa, 0.0337 GPa, 0.0503 GPa, 0.0679 GPa, 0.0830 GPa, 0.100 GPa, 0.115 GPa, 0.150 GPa, 0.230 GPa, 0.399 GPa, 0.807 GPa, 1.0 GPa, 2.5 GPa. The curve corresponding to the critical pressure is shown by a dashed line. Black and open circles show the liquid and vapour densities respectively, from~\cite{Bhatt:2006fw_sm}, with a black star showing their estimate of the critical point. The inset shows the minimum value of the derivative of the curves obtained by Nested Sampling, with the arrow indicating the pressure value where the plateau between vapour and liquid phases diminish at the critical point.}
\label{fig:Al_ns_and_GEMC}
\end{figure}

By calculating the expected density as a function of temperature, the equation-of-state can be calculated using Nested Sampling. These are plotted for the low pressure region together with Gibbs ensemble Monte Carlo simulation results from Ref~\cite{Bhatt:2006fw_sm} in Figure~\ref{fig:Al_ns_and_GEMC}.  
The critical parameters for the different aluminium potentials determined by Nested Sampling are summarised in Table~\ref{table:Al_critical_params}. 

\begin{table}[hb]
\caption{Critical temperature and pressure as estimated from the Nested Sampling calculations for the different potential models. Numbers in parenthesis show GEMC results from Ref.~\cite{Bhatt:2006fw_sm} for comparison.}
\begin{tabular}{lll}
\hline\hline
                                & $T_c$ (K)  & $p_c$ (GPa)   \\  %
\hline
LEA-EAM                & $6810 \pm 170$  & 0.262  \\  %
Mishin-EAM            & $5800 \pm 140$ &  0.054  \\  %
NPB-EAM               & $6100 \pm 150$ ($6299\pm48$)  & 0.083 ($0.0896\pm0.0019$)  \\  %
MD-EAM                 & $3340 \pm 140$ ($3381\pm13$) & 0.042 ($0.044\pm0.002$) \\  %
\hline\hline
\end{tabular}
\label{table:Al_critical_params}
\end{table}

\section{NiTi}

Here we discuss the lowest enthalpy structures we identified for the EAM potential used to describe NiTi~\cite{NiTi_EAM_sm}.
The enthalpic and energetic ordering of these structures is the same for all pressures considered in this paper.
Therefore we shall restrict our discussion to their energy values.

The EAM potential gives rise to a number of low energy structures that show some similarity to the experimentally observed low temperature structure (B19') and to other structures found using DFT (namely B19 and BCO). The lowest energy structure of the EAM potential is orthorhombic, but with significant displacement of the atoms from their high temperature average B2 positions.  Table~\ref{table:niti} below gives the structural parameters of the three structures with the lowest energies. In addition, we found that a multitude of local minima exist between just 10 and 15 meV/atom above the lowest energy structure. The symmetries of these configurations were identified using the {\sc findsym} package~\cite{findsym1_sm}.

\begin{table}
\caption{Low energy structures of NiTi according to the EAM potential. Lattice vectors are in \AA, energies are per atom and are relative to the B2 phase.}
\begin{tabular}{lcccccc}
\hline\hline
Symmetry     		& $a$  & $b$  & $c$   & $\beta$  & Wyckoff positions & Energy (meV)\\  %
\hline
Pm$\bar 3$m (B2)    & 3.0098& 3.0098  & 3.00980  & 90 & Ni: a, Ti: b & 0  \\  %
Pmm2 (B19X) 		&4.02186&4.40909  &3.01484 & 90 & Ni: b (z=-0.04073), c (z=0.04305) Ti: a (z=-0.4434), d (z=0.4458) & -53.8\\
C2/m  (B19)                       &4.46697 &   4.02169& 3.00532 & 98.1 & Ni: a, Ti: d & -51.1\\
P2$_1$/m (BCO)              &2.99360 &    4.00083  &4.88392 & 107.8  & Ni: e (x=0.08755, z=-0.3249) Ti: e (x=0.3583, z=0.2167) & -46.8\\
\hline\hline
\end{tabular}
\label{table:niti}
\end{table}

\end{document}